%% file: orphan.tex
\documentclass[a4paper,useAMS,usenatbib]{mnras}

\usepackage{amsmath,amstext,amsgen,amsbsy,amsopn,amsfonts,theorem}

\usepackage[pdftex]{graphicx}
\usepackage{xspace}
\usepackage{amsmath}
\usepackage{hyperref}
\usepackage{txfonts}
\usepackage{epstopdf}
\usepackage{gensymb}

\usepackage{color,soul}

\makeatletter
\def\mr@ignsp#1 {\ifx\:#1\@empty\else #1\expandafter\mr@ignsp\fi}%
\newcommand{\multiref}[1]{\begingroup%\let\protect\string%
\xdef\mr@no@sparg{\expandafter\mr@ignsp#1 \: }%
\def\mr@comma{}%
\@for\mr@refs:=\mr@no@sparg\do{\mr@comma\def\mr@comma{,}\ref{\mr@refs}}%
\endgroup}
\makeatother

\newcommand{\hypref}[2]{\ifx\href\asklfhas #2\else\href{#1}{#2}\fi}
\newcommand{\Secref}[1]{Section~\multiref{#1}}

\newcommand{\Tabref}[1]{Table~\multiref{#1}}
\newcommand{\tabref}[1]{Tab.~\multiref{#1}}
\newcommand{\Figref}[1]{Figure~\multiref{#1}}
\newcommand{\figref}[1]{Fig.~\multiref{#1}}
\renewcommand{\eqref}[1]{(\multiref{#1})}

\newcommand{\eq}[1]{\begin{align}#1\end{align}}

\usepackage{color}

\definecolor{darkred}{rgb}{0.64, 0.0, 0.0}
\title[Orphan and the LMC]{The total mass of the Large Magellanic Cloud from its perturbation on the Orphan stream}

\input{authors.tex}

\begin{document}

\label{firstpage}

\maketitle

\begin{abstract}
In a companion paper by Koposov et al., RR Lyrae from \textit{Gaia} Data Release 2 are used to
demonstrate that stars in the Orphan stream have velocity vectors significantly misaligned
with the stream track, suggesting that it
has received a large gravitational perturbation from a satellite of the Milky
Way. We argue that such a mismatch cannot arise due to any realistic
static Milky Way potential and then explore the perturbative effects
of the Large Magellanic Cloud (LMC). We find that the LMC can produce precisely the
observed motion-track mismatch and we therefore use the Orphan stream to
measure the mass of the Cloud. We simultaneously fit the Milky Way and LMC potentials and infer that a total LMC mass
of $1.38^{+0.27}_{-0.24} \times10^{11}\,\rm{M_\odot}$ is required to bend the Orphan
Stream, showing for the first time that the LMC has a large and
measurable effect on structures orbiting the Milky Way. This has
far-reaching consequences for any technique which assumes that tracers
are orbiting a static Milky Way. Furthermore, we measure the Milky Way mass within 50 kpc to be $3.80^{+0.14}_{-0.11}\times10^{11} M_\odot$. Finally, we use these results to predict that, due to
the reflex motion of the Milky Way in response to the LMC, the outskirts of the Milky Way's stellar halo should exhibit a bulk, upwards motion. 
\end{abstract}


\section{Introduction} \label{sec:intro}

The total mass of a galaxy is hard to measure
\citep[][]{White2001}. While it can be done in a statistical sense for
a sample of objects using, for example, weak lensing \citep[see
  e.g.][]{Mandelbaum2006} or halo abundance matching
\citep[e.g.][]{Moster2013}, for individual galaxies, the total mass is
always an extrapolation. For high-luminosity galaxies, a variety of
galaxy-weighing methods exist which rely on either the availability of
kinematic tracers \citep[e.g.][]{Wilkinson1999,Cappellari2006,Xue2008}
or the presence of a gravitational lensing signal
\citep[e.g.][]{Kochanek2001}, or both
\citep[e.g.][]{sand02,Treu2004,Auger2010}.
%\CL{ X-ray gas \citep[e.g.][]{buote07} or their full combination \citep[e.g.][]{newman13}}.

In all cases, the measurement is limited by the extent of the
kinematic tracers which rarely reach out to a substantial fraction of
the virial radius. For dwarf galaxies, the uncertainty related to the
extrapolation of the mass probed to the virial radius is exacerbated
by the absence of tracers at intermediate and large distances \citep[e.g.][]{aaronson_1983,kleyna_etal_2005,walker_etal_2009,agnello_evans_2012,errani_etal_2018}.

Stellar streams have recently been demonstrated to provide a new and
independent method to gauge the mass distribution inside our own Milky
Way \citep[see e.g.][]{johnston99,koposov_gd1,mLCS,Kupper2015} and a handful of
nearby galaxies
\citep[e.g.][]{Ibata2004,Fardal2013,Amorisco2015}. Tidal streams not
only measure the mass within the extent of the stream, but also
provide constraints on the slope of the total matter density
\citep[see][]{mLCS}, thus allowing us to pin down the host's mass
further out \citep[see also][]{Bonaca2018}. Moreover, streams are
sensitive probes of the shape of the underlying potential
\citep[see][]{Ibata2001,helmi04,Johnston2005,Fellhauer2006,law_majewski_2010,Bowden2015}
and their twisting and fanning can be used to understand the amount of
asphericity of the Dark Matter halo \citep[see][]{Pearson2015,stray}.

As numerous examples of halo substructure started to be discovered in
the Milky Way, perturbations of stellar streams were put forward
as a new promising diagnostic, capable of constraining masses of
Galactic satellites, even those entirely devoid of light \citep[see
  e.g.][]{Ibata2002,Johnston2002, Siegal2008, Carlberg2009, Yoon2011,ngan_carlberg,
  Erkal2015}. Such perturbations arise as a result of the
stream-subhalo interaction during which the paths of the stars around
the point of the closest approach are slightly altered after receiving
a velocity kick from the passing deflector. The affected stars begin
to re-arrange their orbits, and, with time, small differences can
accumulate to produce observable signatures \citep[e.g.][]{carlberg_pal5, carlberg_gd1, bovy2017, Erkal2017,
  deboer_gd1, PW2018}. However, when the perturbation is strong enough, 
 its effect on the stream can be spotted soon after the
interaction. For fast encounters with dense objects, this can eject stars 
from the stream and lead to spur-like features \citep[e.g.,][]{Bonaca2018spur}.
If the perturber is very massive and less dense, the kick received by the stars can be
large enough to cause the stream's velocity vector to become misaligned
with the original direction of the stream's motion \citep{tuciii_modelling}. %For example, the
%Tucana III stream is predicted to exhibit such a misalignment caused
%by its fly-by near the Large Magellanic Cloud
%\citep[LMC,][]{tuciii_modelling}. Unfortunately, Tuc III's 
%current distance uncertainties have made it difficult to confirm whether
%the stream is misaligned.

With a stellar mass of $2.7\times10^9 \,\rm{M_\odot}$
\citep{vandermarel_2006}, the LMC is our Galaxy's largest satellite
and thus is certainly capable of causing detectable perturbations in
some of the Milky Way stellar streams. The number of streams
affected, and the strength of the effect, depends on the total mass of
the Cloud and its distance. While the latter is well known
\citep[$\sim 50$ kpc,][]{lmc_dist}, the former is currently not
constrained. Efforts to measure the Cloud's mass directly from the
dynamics of the LMC clusters \citep{schommer_etal_1992} or via
modelling of its rotation curve
\citep{vandermarel_kallivayalil_2014} have both given relatively
modest estimates of $\sim2\times10^{10} \,\rm{M_\odot}$ within $\sim 9$
kpc. However, wide-field surveys have recently revealed a wealth of
debris in the outskirts of the LMC
\citep[e.g.][]{mackey_lmc_disk,besla_etal_2016,BK2016,Deason2017,Belokurov2017,mackey_2018,Navarrete2018,choi_etal_2019,nidever_etal_2019},
indicating that the Cloud may be significantly more massive. Indeed,
several distinct lines of reasoning have suggested an LMC mass of
$\sim 1$-$2.5\times10^{11} \,\rm{M_\odot}$. First, requiring that the LMC and
SMC are bound together requires an LMC mass above $\sim 10^{11}
\,\rm{M_\odot}$ \citep{lmc_pms}. Second, accounting for the LMC in the timing
argument between the Milky Way and M31, as well as in the nearby Hubble flow,
gives an LMC mass of $2.5\times10^{11} \,\rm{M_\odot}$
\citep{penarrubia_lmc_mass}. Third, live N-body models of the LMC on a
first infall orbit \citep{laporte2018} favor a massive LMC of
$2.5\times10^{11}\,\rm{M_\odot}$ to explain the shape
of the HI warp \citep{levine06} through the response of the halo
\citep{weinberg98}. Finally, abundance matching
\citep{Moster2013,behroozi_etal_2013} based on the LMC's stellar mass
of $2.7\times10^9 \,\rm{M_\odot}$ \citep{vandermarel_2006} gives a peak halo
mass of $2\times10^{11} \,\rm{M_\odot}$. These results suggest that the LMC's
diminutive size on the sky belies its true mass.

The influence of the LMC on the behavior of stellar streams around the
Milky Way was originally considered in \cite{law_majewski_2010} who
discussed the interaction of a relatively light Cloud ($<6\times
10^{10}\,\rm{M_\odot}$) with the Sagittarius stream and found that it could
have a significant effect. \cite{vera-ciro_helmi_2013} followed this
up and argued that including a $8\times10^{10} \,\rm{M_\odot}$ LMC could
change the Milky Way halo shape inferred by \cite{law_majewski_2010},
making it more spherical. Along these lines, \cite{gomez_etal_2015}
simulated the infall of a $1.8\times10^{11} \,\rm{M_\odot}$ LMC and found
that it would induce a significant reflex motion in the Milky Way
which would affect the Sagittarius stream. More recently,
\cite{tuciii_modelling} studied the effect of the LMC on the Tucana
III stream and found that it could induce a substantial proper motion
perpendicular to the stream which would be detectable in \textit{Gaia}
DR2. They further argued that the size of this offset could be used to
measure the mass of the LMC.

Here we present the results of a comprehensive modeling of the Orphan
Stream \citep[OS,][]{orphan_disc_g,orphan_disc_v}, inspired by its
recent detection in the \textit{Gaia} data presented in \cite{koposov_etal_orphan}. In the \textit{Gaia} Data Release 2
\citep[see][]{Brown2018}, RR Lyrae stars have been used to trace the
OS over $>200^{\circ}$ on the sky, revealing the details of its
behavior in 5 out of 6 dimensions of the phase-space, i.e. on-sky position, distance, and proper motions. The stream's
track is shown to swing by some $~20^{\circ}$ when the 3D
positions of its Southern members are compared to those in the
North. Moreover, across tens of degrees, the proper motions of the
OS's RR Lyrae are demonstrably offset from the direction delineated by
the stream's extent. We propose that the large-scale wobble of the
Orphan's track together with the stream-proper motion misalignment 
reported in \cite{koposov_etal_orphan} are best explained as a result
of an interaction between the stream and the LMC. We explore a range
of Milky Way mass models and allow for the Galaxy's Dark Matter halo to
be aspherical with the flattened axis oriented in an arbitrary
direction. We also include the effects of the Milky Way' response to the
LMC's infall \citep[see e.g.][]{gomez_etal_2015}.

This paper is organized as follows. In \Secref{sec:data} we present
the observed misalignment in the Orphan stream with a new
technique. In \Secref{sec:mw_fit} we attempt to fit the Orphan stream
in the presence of just the Milky Way and show that even with a
flexible potential, it is not possible to get a good match. Next, we
perform the same fits including the LMC in \Secref{sec:lmc_fits} which
can reproduce the Orphan stream. These fits constrain the LMC mass and
the shape of the Milky Way halo. We discuss the meaning and
implications of these results in \Secref{sec:discussion} before
concluding in \Secref{sec:conclusions}.

\begin{figure}
\centering
\includegraphics[width=0.49\textwidth]{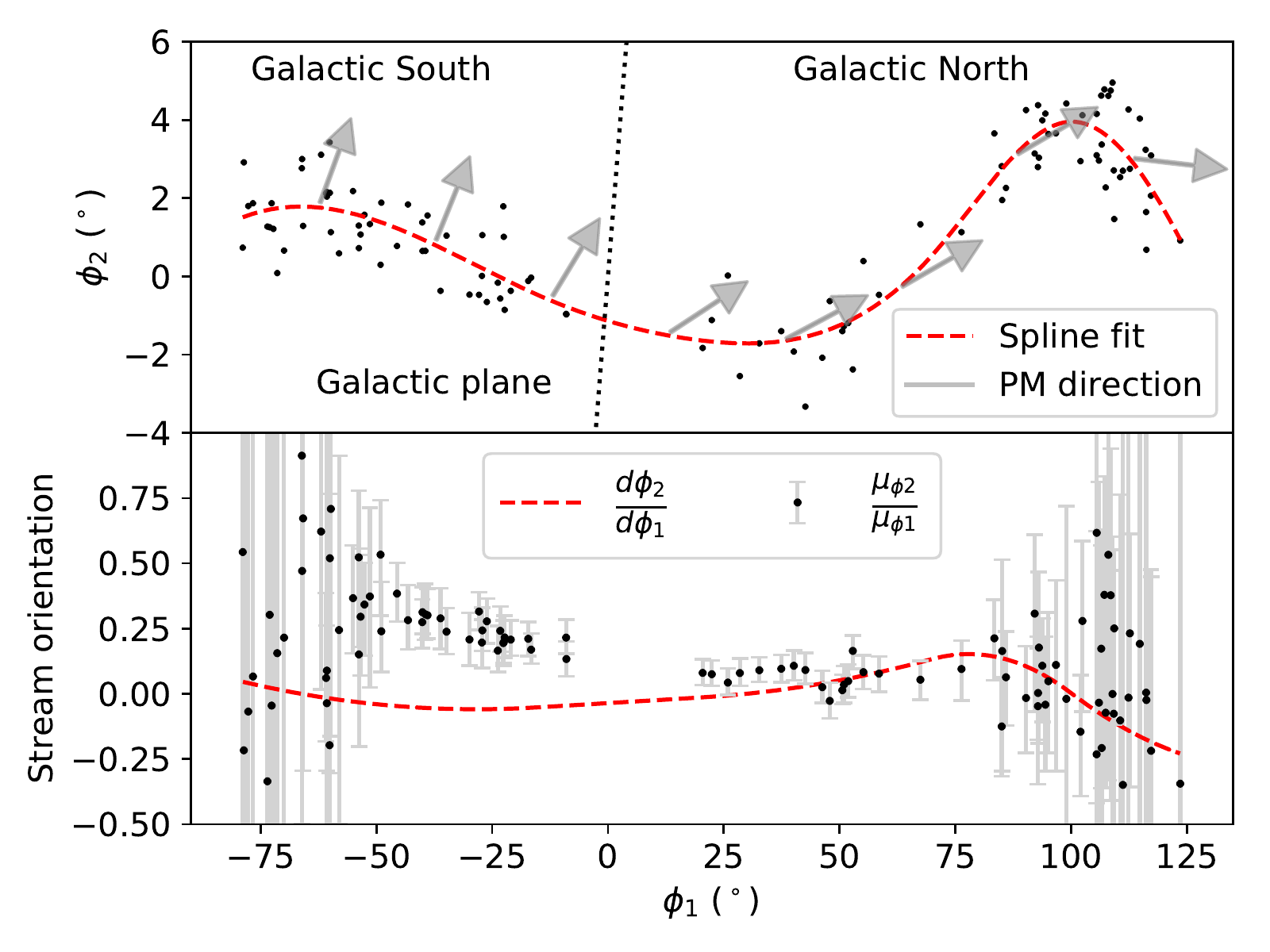}
\caption{Stream track and ratio of proper motions showing misalignment in the Orphan stream. \textbf{Top panel} shows the Orphan stream in coordinates aligned with the stream. The black points are RR Lyrae from \protect\cite{koposov_etal_orphan}, the dashed-red line shows a cubic spline fit to these points, and the dotted line shows the galactic plane. The grey arrows show the reflex-corrected proper motion direction at 25$^\circ$ intervals along the stream. In a stream orbiting a static, non-interacting Milky Way, these arrows would be expected to point along the stream. Note that the final proper motion direction at $\phi_1 \sim 110^\circ$ has a large uncertainty and thus its misalignment is not significant.  \textbf{Lower panel:} Ratio of the reflex corrected proper motions along the stream (black point with grey bars) and the slope of the stream (dashed-red curve). There is a mismatch for $\phi_1 < 50^\circ$ where the stream track has a negative slope but the proper motions indicate a positive slope. This mismatch is the strongest below $\phi_1 \sim 0^\circ$ which corresponds to the Southern Galactic hemisphere.} 
\label{fig:orphan_data}
\end{figure}

\section{Misaligned Orphan stream} \label{sec:data}

Streams can be shown to approximately delineate orbits
\citep{sanders_binney_2013}, which implies that the constituent stars
move mostly along the stream. Indeed, this near-alignment has been
proposed as a way to measure the velocity of the Sun
\citep{majewski_r0lsr_2006,malhan_ibata_reflex,hayes_sgr_lsr} and,
naturally, as a means of finding streams \citep{malhan_stream_finder}. In
order to see how this motion along the stream relates the debris
path and its proper motion, let us consider a stream in an on-sky
coordinate system where it follows a track $(\phi_1(s),\phi_2(s))$
parameterized by $s$. If the proper motions are aligned with the
stream track, then the tangent to the stream, $(\frac{d\phi_1}{ds},
\frac{d\phi_2}{ds})$, should be proportional to the motion of the
stars along the stream, $(\frac{d\phi_1}{dt}, \frac{d\phi_2}{dt}) =
(\mu_{\phi1}, \mu_{\phi2})$. Note that we assume that the proper
motions are corrected for the Solar reflex. We emphasize that
$\mu_{\phi1}$ is the proper motion in $\phi_1$ without the traditional
$\cos(\phi_2)$ correction. In practice, we compare the slope of the
stream on the sky, $\frac{d\phi_2}{d\phi_1}$, to the ratio of the
proper motions in the stream, $\frac{\mu_{\phi2}}{\mu_{\phi1}}$. Replacing
$\phi_2(s)$ with the distance to the stream, $r(s)$, this argument also
implies that $\frac{v_r}{\mu_{\phi 1}}$ can be compared with $\frac{dr}{d\phi_1}$
where $v_r$ is the Solar reflex corrected radial velocity, i.e. $v_{\rm gsr}$. Note
that these comparisons can be made in any coordinates and can be used to
easily determine whether the stream has been significantly perturbed.

Observationally, the motion-track alignment has been demonstrated in
several streams in the Milky Way. For example, both the GD-1
\citep{gd1_disc} and the Palomar 5 streams \citep{pal5_disc} have
proper motions closely aligned with their stream tracks
\citep[][respectively]{koposov_gd1,pal5_pm}.

In \Figref{fig:orphan_data} we investigate whether the assumption of
motion-track alignment holds for the OS, as traced using RR
Lyrae from \textit{Gaia} DR2 \citep[see][]{koposov_etal_orphan}. The $(\phi_1,
\phi_2)$ coordinates are obtained by a rotation of the celestial
equator to a great circle with a pole of $(\alpha_{\rm
  pole},\delta_{\rm pole}) = (72^\circ,-14^\circ)$ and a zero point at
$(\alpha,\delta)=(191.10487^\circ,62.86084^\circ)$. The top panel
shows the RR Lyrae on the sky along with a cubic spline fit
(dashed-red curve). The cubic spline uses fixed nodes with a spacing
of 30$^\circ$. The bottom panel shows the slope of this track
(dashed-red curve) along with the ratio of the reflex-corrected proper
motions (black points with grey error bars). The error bars come from
Monte Carlo sampling of the proper motions and the distances given the
observational uncertainties. The errors are largest at the ends of
the visible stream due to their relatively large distances ($r > 30$ kpc).

The stream track and proper motions are misaligned for $\phi_1 <
50^\circ$ with the strongest mismatch below $\phi_1 <
0^\circ$. Interestingly, this corresponds to the previously unseen
portion of Orphan in the Southern Galactic hemisphere where the stream
is closest to the LMC. To give a sense of the magnitude of the
misalignment, in the top panel light grey arrows show the proper
motion direction averaged in 25$^\circ$ intervals along the
stream. There is a clear misalignment in the South where the proper
motions point upwards but the stream has a gentle negative slope. Note
that the top panel of \Figref{fig:orphan_data} slightly exaggerates
the misalignment due to the aspect ratio of the figure.

\section{Fitting the Northern part of the Stream} \label{sec:mw_fit}

Given the strong misalignment seen in \Figref{fig:orphan_data}, it is
clear that orbit modelling will fail for this particular stream since
the orbit's projection on the sky and its proper motion are always
aligned by construction. Therefore, in what follows we instead use
realistic stream models. We start by fitting only the Northern portion
of the OS in a Galaxy model with an aspherical Dark Matter halo. This
allows us to both compare to the results in the literature (although a
genuine stream model has not yet been used to explain even the
Northern data) and to better elucidate the effect of the LMC. We then
demonstrate how different mass LMCs can deflect the Southern portion
of such a model stream and effortlessly bring it into agreement with
the Orphan data. Later, in \Secref{sec:lmc_fits}, we will explore
models of the entire stream.

\subsection{Setup} \label{sec:mw_setup}

We generate streams using the modified Lagrange Cloud Stripping (mLCS)
technique developed in \citet{mLCS}. This method rapidly generates
streams by ejecting swarms of test particles from the Lagrange points
of a progenitor whose gravitational potential is represented
analytically. We model the Orphan's progenitor as a $10^7 \,\rm{M_\odot}$ \citep[in
approximate agreement with the observationally-motivated mass
estimates from][]{koposov_etal_orphan} Plummer sphere with a scale radius of $1$ kpc. These
parameters were chosen to roughly match the width of the OS. For the Milky Way gravitational potential, we choose a
generalized version of \texttt{MWPotential2014} from \cite{galpy}
which consists of an NFW halo \citep{nfw_1997}, a Miyamoto-Nagai disk
\citep{mn_disk}, and a power-law bulge with an exponential cutoff. We
parameterize the NFW halo in terms of its mass, $M_{\rm NFW}$, a scale
radius, $r_{s, \, \rm{NFW}}$, and a fixed concentration $c=15.3$. The
Miyamoto-Nagai disk has a mass of $6.8\times10^{10} \,\rm{M_\odot}$, a scale
radius of $3$ kpc, and a scale height of $280$ pc. For simplicity, we
replace the bulge with a Hernquist profile \citep{hernquist_profile}
with the same mass ($5\times10^9 M_\odot$) and a scale radius of 500 pc. For the fits, we keep
the disk and bulge fixed but allow the mass of the NFW halo to
vary. We introduce a flattening in the halo potential by writing the
NFW potential as \eq{\phi_{\rm NFW}(x,y,z) = -\frac{GM_{\rm
      NFW}}{\tilde{r}} \frac{\log(1+\frac{\tilde{r}}{r_s})}{\log(1+c)
    - \frac{c}{1+c} } , \label{eq:flat_NFW}}
where $M_{\rm NFW}$, $r_s$, and $c$ are the NFW halo's mass, scale
radius, and concentration respectively, while
\eq{ \tilde{r}^2 = x^2+y^2+z^2 + (\frac{1}{q^2}-1) (\mathbf{\hat{n}}\cdot \mathbf{x})^2 , }
$\mathbf{\hat{n}}$ is the unit vector in which the potential is
flattened by $q$, and $\mathbf{x} = (x,y,z)$. For $\mathbf{\hat{n}}$
pointed in the $z$ direction, this reduces to the commonly used form \citep[e.g.][]{richstone1980,evans1994}. Note that that our potential (\ref{eq:flat_NFW}) has the advantage that the force components are analytic, thus speeding our computations. However, it has the disadvantage that the density is not everywhere positive (unless $q =1$). This defect normally occurs at radii well beyond the regime relevant for our simulations. It is also worth bearing in mind that the density contours can deviate substantially from an ellipsoidal shape, so $q$ is best interpreted as a diagnostic of the effect of the LMC on the OS orbit rather than the intrinsic flattening of the Milky Way halo. We will discuss this interpretation further in \Secref{sec:mw_halo_shape}.

\begin{figure*}
\centering
\includegraphics[width=0.95\textwidth]{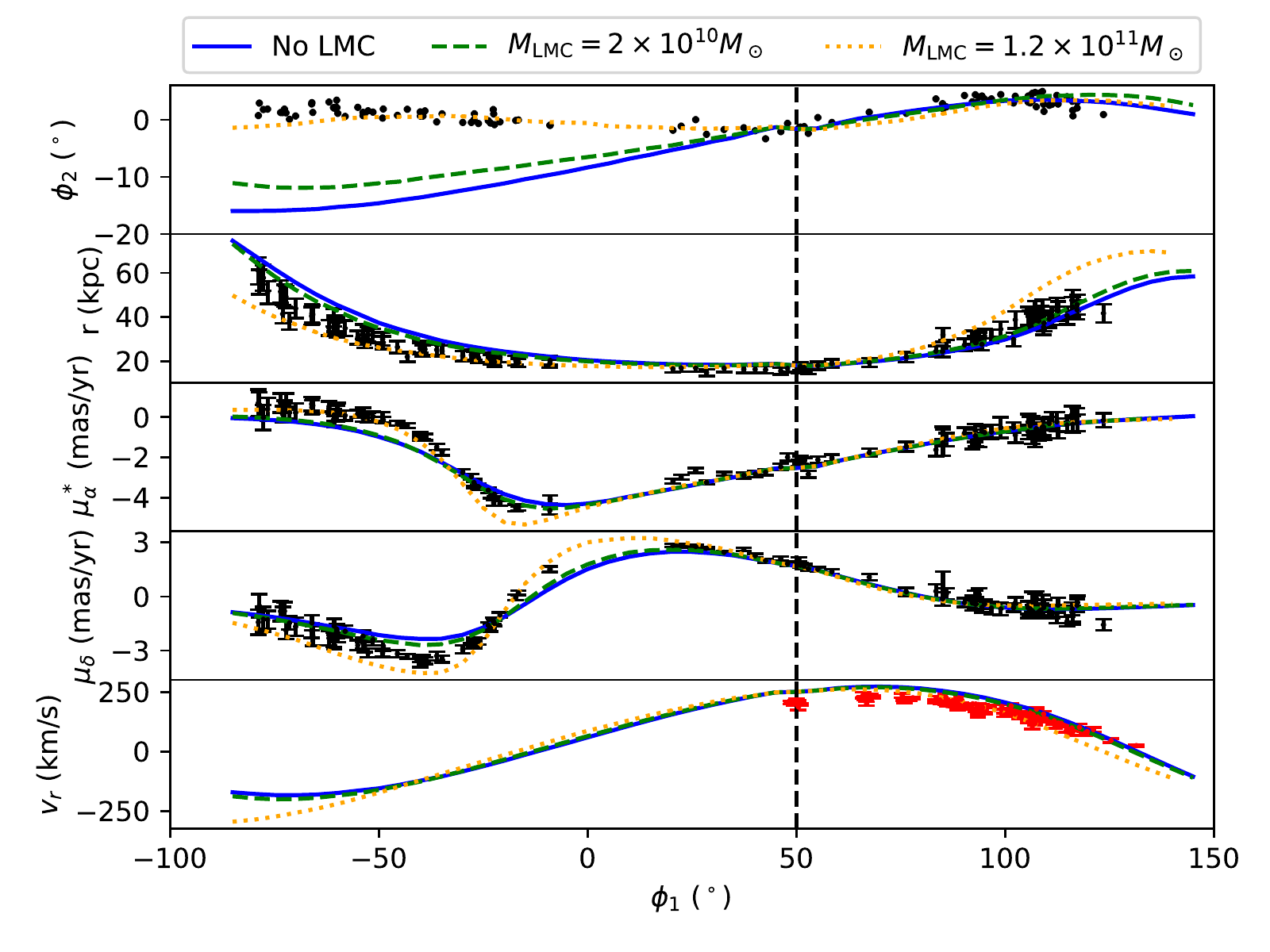}
\caption{Best fit to the Northern portion of the OS in the Milky Way. From top to bottom the panels show the stream on the sky, the run of distances along the stream, the proper motion in RA, the proper motion in Dec, and the radial velocity. In the top four panels, the black points show the RR Lyrae from \protect\cite{koposov_etal_orphan} and the red point shows an RR Lyra which was removed from the fit since it appeared to be an outlier. The red points with error bars in the bottom panel show the radial velocities which are not used in the fit. The blue line shows the best-fit to the Northern portion of the OS ($\phi_1 > 50^\circ$, marked by the dotted line) in a Milky Way with an oblate halo. The dashed-green, dotted-orange lines show the same best-fit (i.e. same exact parameters) with a $2\times10^{10} \,\rm{M_\odot}$, $1.2\times10^{11} \,\rm{M_\odot}$ LMC respectively. This shows that the LMC should have a significant effect on the OS and that, at least in this simple case, a $1.2\times10^{11}\,\rm{M_\odot}$ LMC broadly matches the observations. We stress that streams do not represent best-fits but rather are meant to showcase the effect of the LMC. Interestingly, it shows that the LMC also has a substantial effect on the Northern portion of the OS (e.g. large change in the distance) and thus fits to the Northern portion of the OS which neglect the LMC will likely be biased. } 
\label{fig:mw_fit}
\end{figure*}

We take the Sun's velocity relative to the Galactic standard of rest
to be $(11.1,245,7.3)$ km/s. The $y$ component of this velocity is
based on the proper motion of Sgr A* \citep{pm_sgr_a} and the distance
to the Galactic center of 8.1 kpc from \cite{s2_GR_R0}. The $x$ and
$z$ components come from \cite{schonrich_vlsr}. We note that since the OS mostly probes the outer part of the Milky Way, we will
keep these parameters fixed instead of relating the circular velocity
at the Sun's location to the potential.

We fit the RR Lyrae from \cite{koposov_etal_orphan}. These data are shown as
black points in \Figref{fig:mw_fit}. In order to clarify the effect of
the LMC, in this Section we choose to only fit the stars with $\phi_1
> 50^\circ$. These are also the stars which do not show any
significant proper motion offset (see \figref{fig:orphan_data}) so we
expect these have not been heavily perturbed by the LMC. This $\phi_1$
range also corresponds to the portion of the stream with previously
available data which were modelled in \cite{newberg_etal_2010} and
\cite{hendel_etal_2018} so we can also compare our results to those in
the literature. We note that for a sub-section of the stream, \cite{koposov_etal_orphan} also presented two additional debris track measurements,
one based on Red Giant Branch (RGB) stars from \textit{Gaia} DR2 and
one computed using the RGB/Main Sequence Turn-Off stars in DECaLS
\citep{DECALS2018}. These tracks show good agreement overall, though 
a small offset near $\phi_1 \sim 100^\circ$ (see their Fig. 8) is
reported. For the sake of consistency, we choose to only fit the GDR2
RR Lyrae (covering the entire detected length of the stream). We will
discuss possible implications of our choice in \Secref{sec:discussion}.
\cite{koposov_etal_orphan} also present radial velocities of Orphan stream
stars from SDSS which we compare against our best-fit streams but 
we do not use in our fits.

Since the progenitor of the OS is not known \cite[but see the
discussion of a possible association between the OS and the Gru 2
satellite in][]{koposov_etal_orphan}, we should in principle include its
position as a free parameter in our model. Indeed, in
\Secref{sec:lmc_fits} we consider multiple progenitor positions and
find that our results are independent of this choice. However, in this
section we choose to place the progenitor at $\phi_1 = 50^\circ$ since
we will only be fitting the RR Lyrae with $\phi_1 > 50^\circ$. This
forces the progenitor to be on the stream so that we can later include
the LMC and see how it deflects the stream. To describe the
progenitor's present day position and velocity we then have 5
additional parameters: the two proper motions $(\mu_{\alpha,\,\rm
  prog}^*,\mu_{\delta,\,\rm prog})$, the radial velocity ($v_{r,\,\rm
  prog}$), the distance ($d_{\rm prog}$), and the latitude ($\phi_{\rm
  2,\,prog}$). Once these parameters are specified, we then evolve the
stream backwards for 5 Gyr and then use the mLCS technique to generate
the stream's present configuration. We strip stars with a Gaussian
spread of $100$ Myr around each pericenter with respect to the Milky
Way. In order to ensure that there are enough particles in the stream,
we strip 5,000 stars per pericentric passage.

\subsection{Likelihood}

The likelihood of the set of RR Lyrae from \cite{koposov_etal_orphan} is
computed as follows. For each star $i$, four components of its
phase-space position $m_i$ are modelled, namely: latitude
$\phi_2$, heliocentric distance, and both proper motion
components. As described above, in this section we only fit the RR
Lyrae with $\phi_1 > 50^\circ$. For each RR Lyra and each observable,
we fit a straight line to the $m(\phi_1)$ distribution of particles in
the simulated stream within $2.5^\circ$ in $\phi_1$ of the RR
Lyrae. This fit returns the maximum likelihood value of the observable
at the location of the RR Lyra, $m_{i,\,\rm sim}$, and the width of
that observable, $\sigma_{i,\,\rm sim}$. The log likelihood takes the
form

\eq{ \log\mathcal{L}_i = -\frac{1}{2} \log \left( 2\pi (\sigma_{i,\,\rm obs}^2 + \sigma_{i,\,\rm sim}^2 )\right) - \frac{1}{2} \frac{(m_{i,\,\rm obs}-m_{i,\,\rm sim})^2}{\sigma_{i,\,\rm obs}^2 + \sigma_{i,\,\rm sim}^2} ,}
where $\sigma_{i,\,\rm obs}$ is the observed uncertainty for each RR
Lyrae and $m_{i,\,\rm obs}$ is the value of the observable for $i$th
star. This is then summed over all of the RR Lyrae and all of the
observables. Note that we ignore the covariance between the two proper
motions in our likelihood. This is justified since the median of the absolute magnitude of the correlation is small,
0.22 for the entire sample. If we restrict to the range where the proper motion
errors are small, $-50^\circ < \phi_1 < 75^\circ$, the median drops to 0.14.

\subsection{MCMC setup}

We perform the Markov Chain Monte Carlos using \textsc{emcee}
\citep{emcee}. Our model has 10 free parameters: 5 that describe the
6d position of the OS progenitor, the mass and scale radius
of the NFW halo ($r_s, M_{\rm NFW}$), as well as its flattening
($q_{\rm NFW}$) and the orientation of the major axis on the sky
($l_{\rm NFW},b_{\rm NFW}$). We use uniform priors over a broad range
for all of these variables which are listed in \Tabref{tab:priors}. We use 200 walkers
for 1500 steps with a burn-in of 750 steps.

\subsection{Stream fit in an aspherical Milky Way} \label{sec:mw_fit_subsec}

With the setup now defined, we proceed to fit the OS. Before beginning the MCMC sampling, we use the downhill
simplex method to find the best-fit stream. We randomly sample 100
points from our priors in \Tabref{tab:priors} and find two distinct
solutions. Namely, we find that both a prolate and an oblate halo
(with different orientations) can fit the Northern part of the OS. We will return to these two solutions in \Secref{sec:lmc_fits}
but in this section, for simplicity, we restrict the halo to be
oblate.

The best-fit stream is shown with a blue curve in
\Figref{fig:mw_fit}. By design, this model provides a good fit to
the Northern OS data, i.e. for stars with $\phi_1 > 50^\circ$. Outside of
this range, the model matches reasonably well the distance and proper
motions of the OS, but it shows a significant deviation in
the track on the sky. The inability to simultaneously reproduce the
proper motions and the stream track was foreshadowed in
\Figref{fig:orphan_data} where we showed that the proper motions and
the stream track had a pronounced offset. Note that we have also
attempted to fit the entire range of the OS data in the
presence of the Milky Way but we are unable to simultaneously match
all observables (see next Section).

Interestingly, if we take this best-fit model from \Figref{fig:mw_fit}
(i.e. the derived phase-space coordinates of the progenitor together
with the potential parameters) and include a $2\times10^{10} \,\rm{M_\odot}$
LMC (for the details of the Cloud's orbit see
Section~\ref{sec:lmc_setup}), we see that the predicted on-sky track
(green dashed line) immediately swings much closer to the observed
track. This shows that the LMC will have a substantial effect on the OS even if we limit its mass to that contained in its inner
regions. If we then increase the LMC mass to
$1.2\times10^{11}\,\rm{M_\odot}$, we obtain a reasonable match to all of the
stream's observables (dotted orange line). This sensitivity suggests
that the OS can be used to weigh the LMC. Let us stress
that the dashed green line and the dotted orange line in
\Figref{fig:mw_fit} are not independently derived models of the
data. These phase-space tracks are obtained by taking the best-fit
Milky Way-only model of the Northern portion of the OS and adding the
LMC-like perturber on an appropriate orbit.

In order to compare our results with the work of \cite{newberg_etal_2010} and
\cite{hendel_etal_2018} who performed orbit fits to Orphan, we compute the
mass enclosed of the Milky Way within 60 kpc. Within this radius we measure a mass of $4.8\pm0.6 \times 10^{11} \,\rm{M_\odot}$ where
we have ignored the flattening of the halo. 
This mass matches well with that of \cite{hendel_etal_2018} who measured a mass of $5.6^{+1.1}_{-1.2}\times10^{11} \,\rm{M_\odot}$. 
However, it disagrees with the results of \cite{newberg_etal_2010} who found a mass of $2.7\times10^{11} \,\rm{M_\odot}$. Our result has an improved precision 
over previous results, expected given the quality
of \textit{Gaia} DR2. 

Before proceeding to actually infer the LMC mass, we note that
\Figref{fig:mw_fit} also shows that while the OS is
strongly deflected by the LMC in the South ($\phi_1 < 0^\circ$), it is
also perturbed by it in the North. This is most evident in the
distances and the stream track on the sky, as indicated by a
noticeable deviation between the dotted orange line and the solid blue
line at $\phi_1>50^{\circ}$. Thus, even if we only fit the OS data
in the North, we would not recover the true Milky Way
potential. Instead, the result of such a model would be biased since
the potential would need to compensate for the influence of the
LMC. Therefore, the effect of the LMC should be considered for all
streams in the Milky Way when attempting to measure the potential.
\begin{table}
\begin{centering}
\begin{tabular}{|c|c|c|}
\hline
Parameter & Prior & Range \\
\hline Orphan & & \\
\hline
$\mu_{\alpha,\, \rm prog}^*$ & Uniform & ($-6,0$) mas/yr \\
$\mu_{\delta,\, \rm prog}$ & Uniform & $(0,6)$ mas/yr \\
$v_{r,\, \rm prog}$ & Uniform & ($-$250,250) km/s \\
$d_{\rm prog}$ & Uniform & (15,20) kpc \\
$\phi_{\rm 2,\,prog}$ & Uniform & $(-10^\circ,10^\circ)$ \\
\hline Milky Way & & \\
\hline $M_{\rm NFW}$ & Uniform & $(6,25)\times10^{11} \,\rm{M_\odot}$ \\
$ r_s$ & Uniform & $(10,30) $ kpc \\
$q_{\rm NFW}$ & Uniform & (0.7,1) or (1,1.3) \\
$l_{\rm NFW}$ & Uniform & $(0^\circ,360^\circ)$ \\
$b_{\rm NFW}$ & Uniform & $(-90^\circ,90^\circ)$ \\
\hline LMC & & \\
\hline $M_{\rm LMC}$ & Log-Uniform & $(10^8,3\times10^{11})\, \,\rm{M_\odot}$ \\
$\mu_{\alpha,\, \rm LMC}^*$ & Normal & $1.91\pm0.02$ mas/yr \\
$\mu_{\delta,\, \rm LMC}$ & Normal & $0.229\pm0.047$ mas/yr \\
$v_{r,\, \rm LMC}$ & Normal & $262.2\pm3.4$ km/s \\
$d_{\rm LMC}$ & Normal & $49.97 \pm 1.126$ kpc
 \\ \hline
\end{tabular}
\caption{Priors for our MCMC fits. Note that the LMC priors are only used in the fits which include the LMC. }
\label{tab:priors}
\end{centering}
\end{table}

\section{Fitting the entire stream} \label{sec:lmc_fits}

As argued in \Secref{sec:intro}, the LMC is the largest satellite of
the Milky Way and is expected to have a substantial mass. Indeed, in
\Secref{sec:mw_fit_subsec} we saw that including the LMC for the
best-fit (of the Northern portion) of the OS in the Milky
Way potential resulted in a large deflection of the
stream. Interestingly, this deflection brought the Southern portion of
Orphan into agreement with the data. In this section we will fit the
entire OS data in the presence of the LMC and we will simultaneously
constrain the LMC mass and the Milky Way halo.

\subsection{Setup} \label{sec:lmc_setup}

The setup is very similar to that described in \Secref{sec:mw_setup}
except that we now add in the LMC. For the LMC's present day position
and velocity, we use proper motions of $\mu^*_\alpha = 1.91 \pm 0.02$
mas/yr, $\mu_\delta = 0.229 \pm 0.047$ mas/yr \citep{lmc_pms}, a
radial velocity of $v_r = 262.2 \pm 3.4$ km/s \citep{lmc_vr}, and a
distance of $49.97 \pm 1.126$ kpc \citep{lmc_dist}. The LMC is modeled
as a Hernquist profile whose mass is left free. For each mass, we fix
the scale radius so that the mass enclosed at 8.7 kpc matches the
measured value of $1.7\times10^{10} \,\rm{M_\odot}$ from
\cite{vandermarel_kallivayalil_2014}. For LMC masses below $2\times
10^{10}\,\rm{M_\odot}$, we set the scale radius to 0.73 kpc. In order to
correctly model the LMC's orbit around the Milky Way, we include
dynamical friction following the results of \cite{jethwa_lmc_sats}. We
choose to place the OS progenitor at $\phi_1 = 6.34^\circ$
which is roughly half way between the two RR Lyrae closest to the
origin. This choice is made to avoid having the progenitor within the
observed portion of the stream. We also consider models where the
progenitor is located at $\phi_1 = -90^\circ$ and $\phi_1 =
130^\circ$. However, since the results are independent of this choice 
we only show results for $\phi_1 = 6.34^\circ$. 

When including the LMC, the stream generation proceeds
almost identically to what is described in \Secref{sec:mw_setup}
except that now the stream's progenitor is rewound and then disrupted
in the combined presence of the Milky Way and the LMC. Our MCMC setup
thus looks similar to the setup described in \Secref{sec:mw_setup}
except that we have 5 additional parameters for the LMC: its mass,
proper motions, radial velocity, and distance. The mass has a
log-uniform prior between $10^8-3\times10^{11} \,\rm{M_\odot}$ and the
observables have Gaussian priors given from their measurements
above. The proper motions, radial velocity, and distance have priors given
by existing observations. \Tabref{tab:priors} contains a list of all the parameters and
their priors.

\begin{figure*}
\centering
\includegraphics[width=0.95\textwidth]{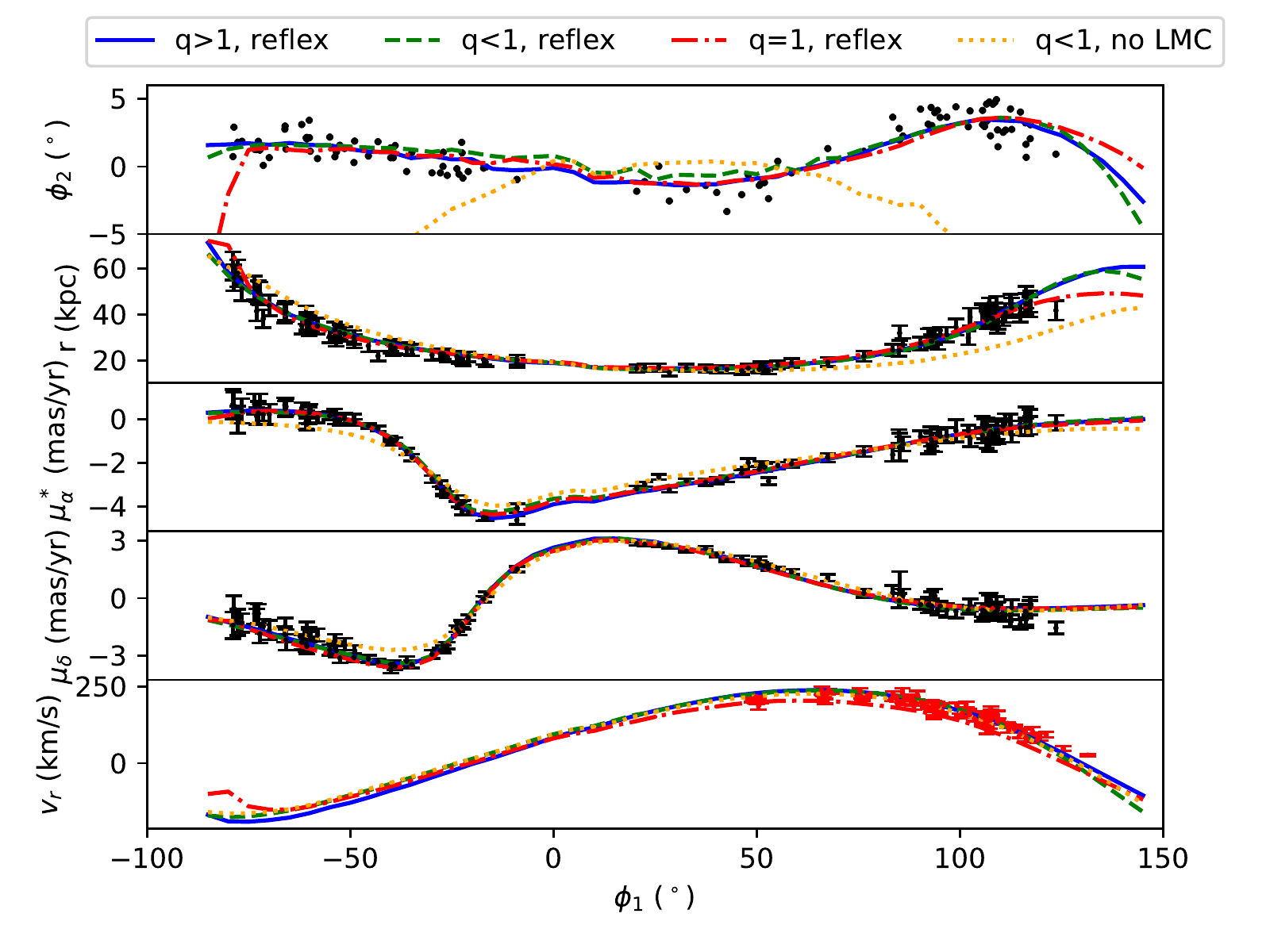}
\caption{Comparison of the best fit models when fitting all of the RR Lyrae. The black points in the top four panels show the observed RR Lyrae from \protect\cite{koposov_etal_orphan}. The red points with error bars in the bottom panel show radial velocities from SDSS which are not included in the fit. The curves show the tracks of the best-fit streams in a prolate halo including the LMC (solid blue line) and a spherical halo including the LMC (dashed green line). In both these cases, the Milky Way halo is represented by a particle which can respond to the LMC. For comparison, a best-fit in an oblate halo without the LMC is included. While this can roughly match most of the observables, it fails to reproduce the stream track on the sky. } 
\label{fig:model_comparison}
\end{figure*}

Given that we now include the LMC, it is also natural to ask whether
we should include the Small Magellanic Cloud
(SMC). \cite{stanimirovic_etal_2004} measured the rotation curve of
the SMC in HI and found a mass of $2.4\times 10^9 \,\rm{M_\odot}$ within 3
kpc. Since this is just the mass within a small aperture, the total
pre-infall mass of the SMC was likely much higher. However, attempts
to model the Magellanic stream seen in HI
\citep{putman_magellanic_stream} have found that the SMC has likely
had multiple pericentric passages with the LMC
\citep[e.g.][]{besla_etal_2012}. Thus, the majority of its mass has
likely been tidally stripped and is now orbiting the LMC. This is
supported by i) the recent proper motion measurements within the SMC
which suggest that the outer portions of the SMC are tidally
disrupting \citep{zivick_etal_2018} and ii) the recent detection of
the SMC's stellar tidal tails
\citep[see][]{Belokurov2017,mackey_2018,CloudsArms}. Therefore, we
choose to ignore the SMC in our analysis but note that a
fraction of the mass we attribute to the LMC is due to the
debris of the SMC.

\subsection{Fitting Orphan in the presence of the LMC} \label{sec:lmc_fits_subsec}

The addition of the LMC to the model allows us to get a significantly
better fit to the OS data than in the Milky Way alone. As
discussed above, this is driven by the fact it is impossible to
produce any considerable motion-track misalignment observed in the OS
without including a massive perturber. Nonetheless, to ensure that our
inference of the LMC mass is not biased by our choice of the Galaxy
model we explore a generously wide range of Milky Way DM halo shapes.  More
precisely, when including the LMC, we consider three separate subsets
of the Milky Way halo potential described in \Secref{sec:mw_setup}: a
spherical halo ($q_{\rm NFW}=1$), an oblate halo ($q_{\rm NFW} < 1$),
and a prolate halo ($q_{\rm NFW} > 1$). The flattening direction of
the halo is only included for non-spherical haloes. In addition to the
halo shape, we also consider a Milky Way which is static and one which
can respond to the LMC's infall. This is important since if the LMC is
massive enough, it will induce a substantial reflex motion in the
Milky Way \citep{weinberg89,gomez_etal_2015}. In order to account for
this shift in the center of mass, we treat the Milky Way as a movable particle
that sources a potential. We give the results of these six fits
in \Tabref{tab:posteriors}. In general we find that the fits in a
reflexive Milky Way halo are best for each choice of the halo shape
and consequently, throughout the rest of this work, we only show the results
for the case where the two galaxies are allowed to move freely.

\begin{table*}
\begin{centering}
\begin{tabular}{|c|c|c|c|c|c|c|}
\hline
Parameter                                           &  sph. MW$+$LMC    & obl. MW$+$LMC        & pro. MW$+$LMC     & sph. rMW$+$LMC  & obl. rMW$+$LMC  & pro. rMW$+$LMC \\\hline 
$M_{\rm NFW}$ ($10^{11} \,\rm{M_\odot}$) &  $ 13.1^{+5.0}_{-3.9}$  & $13.2^{+3.0}_{-2.3}$  & $12.7^{+3.0}_{-2.4}$     & $7.7_{-1.1}^{+1.0}$                         &   $11.3^{+2.2}_{-2.3}$       & $9.4^{+1.5}_{-1.0}$\\
$r_{s}$ (kpc)                                        &  $24.1^{+7.1}_{-6.0}$          & $22.3^{+4.0}_{-3.1}$         & $21.8^{+3.5}_{-2.9}$     & $12.7^{+2.1}_{-2.2}$         &   $21.9^{+3.1}_{-3.8}$     & $17.5^{+2.2}_{-1.8}$\\ 
$q_{\rm NFW}$                                    &  $-$                                   & $0.89^{+0.02}_{-0.03}$ & $1.19^{+0.03}_{-0.03}$ & $-$                                           &    $0.87^{+0.04}_{-0.04}$  & $1.20^{+0.04}_{-0.03}$\\ 
$l_{\rm NFW}$ ($^\circ)$                      &  $-$                                  & $17.5^{+8.4}_{-11.9}$      & $94.2^{+6.9}_{-7.5}$     & $-$                                         &    $-3.8^{+17.5}_{-34.2}$   & $95.5^{+6.3}_{-8.6}$\\ 
$b_{\rm NFW}$ ($^\circ$)                     &  $-$                                  & $-4.0^{+15.8}_{-13.6}$   & $37.2^{+8.1}_{-6.8}$     & $-$                                          &    $13.1^{+13.9}_{-15.1}$   & $32.0^{+5.4}_{-4.5}$\\  
$M_{\rm LMC}$   ($10^{11} \,\rm{M_\odot}$) &  $1.26^{+0.25}_{-0.24}$   & $1.13^{+0.22}_{-0.20}$   & $1.02^{+0.16}_{-0.13}$ & $1.49^{+0.28}_{-0.24}$         &   $1.41^{+0.35}_{-0.33}$   & $1.38^{+0.27}_{-0.24}$ \\ \hline
$\Delta \log \mathcal{L}$                      &   -81                                & -43                                  & -13                                  & -47                                        &             -25                              &  0 \\ \hline
$M_{\rm MW}$(50 kpc) ($10^{11} \,\rm{M_\odot}$) & $4.18^{+0.22}_{-0.23}$  & $4.18^{+0.19}_{-0.25}$ & $4.13^{+0.23}_{-0.25}$ & $4.04^{+ 0.10}_{-0.10}$  &   $3.74^{+0.19}_{-0.22}$    &  $3.80^{+ 0.14}_{-0.11}$     
\end{tabular}
\caption{Posteriors on our Milky Way and LMC properties. We give the posteriors for 6 different setups which include the LMC. For each parameter, we give the median with $1\sigma$ uncertainties. For reference, the $\Delta \log \mathcal{L}$ for our best fit without the LMC shown in \protect\figref{fig:model_comparison} is -457. We do not give any parameters for fits without the LMC since these fits are so poor that the result is not meaningful. The first 3 columns show the results of fits with a spherical, oblate, and prolate halo with a fixed Milky Way which cannot respond to the LMC. The next 3 columns show fits with reflexive models where the Milky Way can move in response to the LMC. These models are denoted as rMW.  The best likelihood for the reflexive, prolate halo. Interestingly, the fits with reflexive haloes require a higher LMC mass of $\sim1.4\times10^{11} \,\rm{M_\odot}$. For reference, the LMC's orbital plane corresponds to ($l_{\rm NFW},b_{\rm NFW}$) = ($-4.5^\circ,5.4^\circ$), the orientation of the best-fit halo in \protect\cite{law_majewski_2010} corresponds to ($l_{\rm NFW},b_{\rm NFW}$) = ($7^\circ,0^\circ$), and the current orientation of the LMC corresponds to ($l_{\rm NFW},b_{\rm NFW}$) = ($89.1^\circ,33.3^\circ$). Curiously, these directions are close to the best-fit halo orientations. The maximum-likelihood values for each of these setups are given in \protect\tabref{tab:mlhood}. }
\label{tab:posteriors}
\end{centering}
\end{table*}
\begin{figure*}
\centering 
\includegraphics[width=8.7cm]{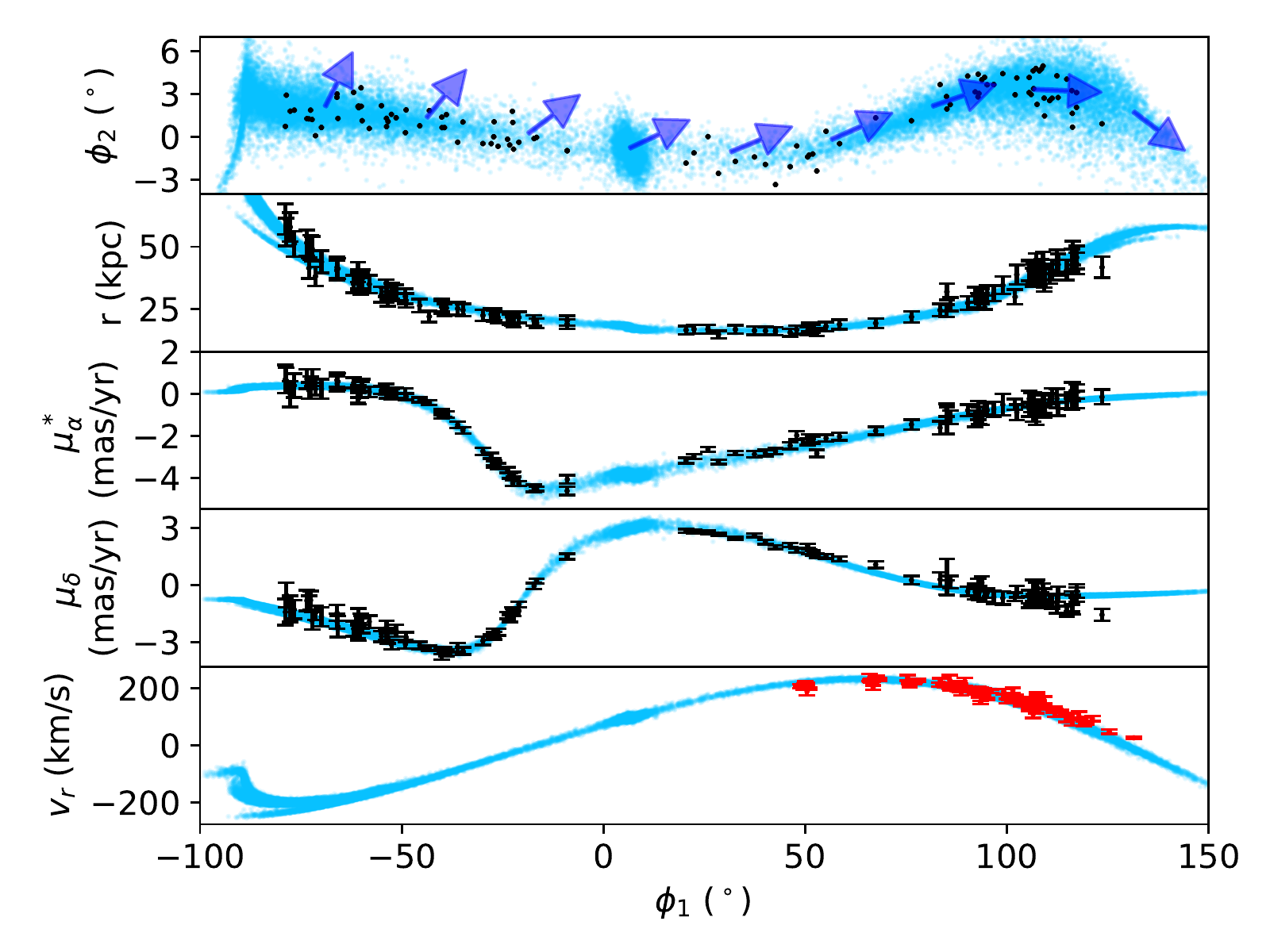}\quad
\includegraphics[width=8.7cm]{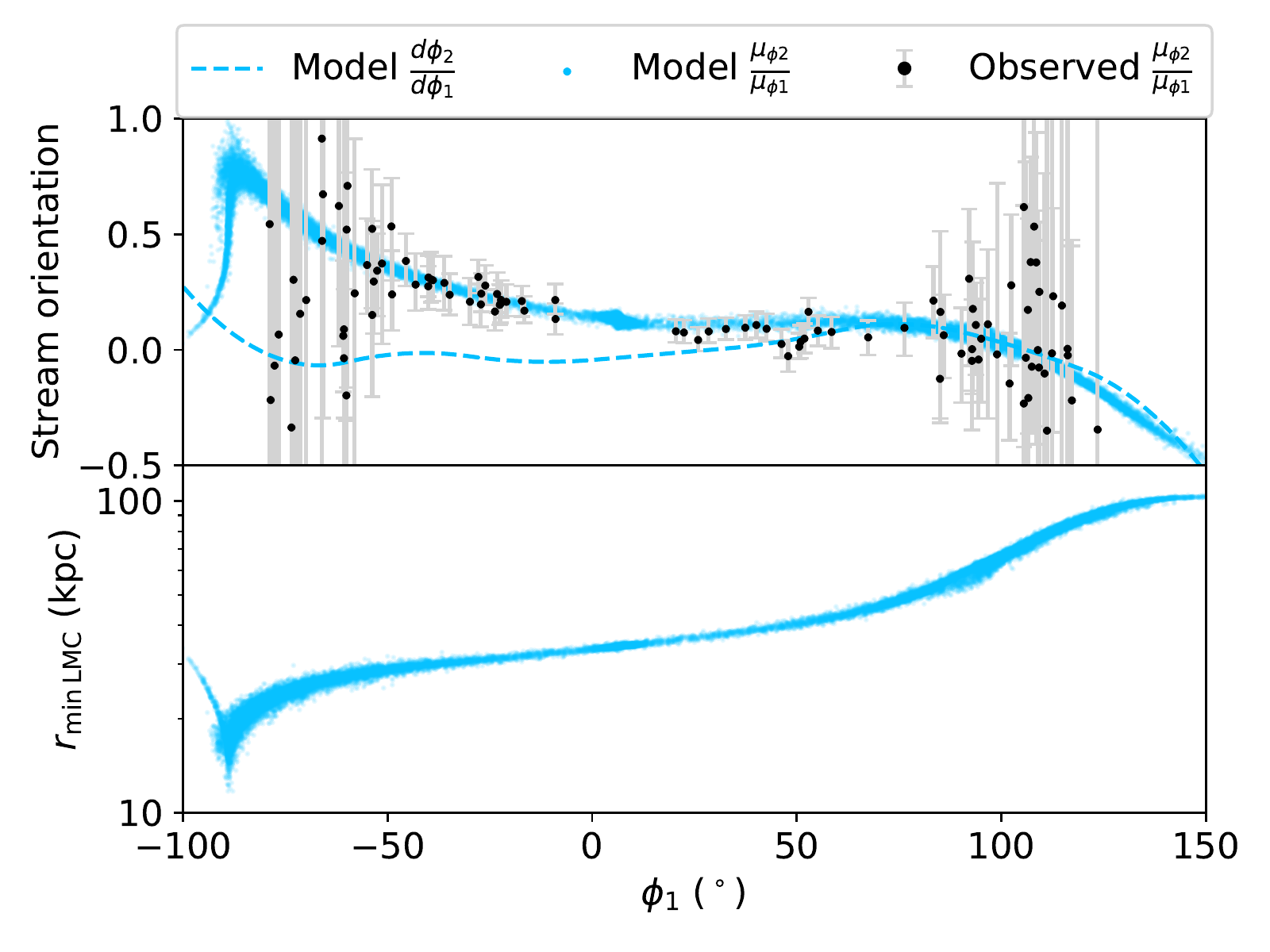}
\caption{Best-fit stream in a prolate Milky Way halo. \textbf{Left panel:} From top to bottom, the panels show the stream on the sky (in coordinates aligned with the stream), heliocentric distance to the stream, proper motion in RA, proper motion in Dec, and radial velocity of the stream. The black points show the RR Lyrae from \textit{Gaia} DR2. The red points with error bars in the bottom panel show the radial velocity of stars in the stream. Note that these radial velocities were not used in the fits. \textbf{Right panel:} In the top panel, the blue points show the ratio of reflex corrected proper motions in the best-fit stream, the black points with error bars show the same ratio for the RR Lyrae, and the dashed-blue line show the slope of the best-fit stream track. The model clearly reproduces the misaligned proper motions seen in the observed stream (see \protect\figref{fig:orphan_data}). The bottom panel shows the closest approach distance to the LMC versus angle along the stream. The left-most portion of the stream has a much more intimate interaction with the LMC which explains why the proper motion offset is largest there. Thus, the misalignment in the OS can be naturally explained by including the LMC.}
\label{fig:fixed_MW_bestfit}
\end{figure*}

The best-fit stream tracks for the
three halo shapes obtained in the presence of the LMC as well as one
stream model with the LMC excluded are shown in \Figref{fig:model_comparison}. 
As is obvious from the Figure, the
addition of the LMC allows us to adequately match the overall
properties of the OS across the entire sky. Some minor
discrepancies still exists, mainly in the behavior of the stream on
the sky (top panel). It is also clear that a spherical Milky Way
provides a poor match to the OS data, especially the oscillation in
the stream track around $\phi_1 \sim 100^\circ$. Allowing the halo to
be axisymmetric (either prolate or oblate) brings the model into a
better agreement with the data. For reference we also show the
best-fit OS model in the Milky Way without the LMC (dotted orange
line). This fit was carried out using a setup identical to that
described in \Secref{sec:mw_fit} except that we place the progenitor at
$\phi_1 = 6.34^\circ$ and fit the entire data range. While the
Milky Way-only model does a reasonable job for some of the observables for a range
of $\phi_1$, it fails miserably in predicting the positions of the OS
debris on the sky.

In \Figref{fig:fixed_MW_bestfit}, we present the stream particles in the
best-fit stream model in a prolate, reflexive Milky Way potential. The left panel
shows the stream observables, which are all a close match to the sample
of RR Lyrae stars from \textit{Gaia} DR2. The top right panel compares
the ratio of the reflex corrected proper motions and the stream track,
mimicking the presentation in \Figref{fig:orphan_data}. We see that
the best-fit model has the same misalignment as the observed
stream. The bottom right panel gives the closest approach distance to
the LMC for each particle in the stream (note the logarithmic scale of
the $y$-axis). We see that the trend is nearly monotonic with stream
particles with smaller $\phi_1$ experiencing a closer approach and therefore a stronger interaction
with the LMC than those with larger $\phi_1$. This explains why the
largest misalignments are seen for $\phi_1<0^\circ$ since this is
where the effect of the LMC is the strongest. These closest approaches are not simultaneous but happen
over a range of times from 350 Myr ago ($\phi_1 \sim -90^\circ$) to 100 Myr ago ($\phi_1 \sim 0^\circ$). 
Thus, we see that the perturbation from the LMC can naturally explain the misalignment seen
in the OS. This best-fit stream and its orbit are publicly available \protect\href{https://zenodo.org/record/2661108\#.XNRIk9NKhXh}{here}.

\begin{figure}
\centering
\includegraphics[width=0.49\textwidth]{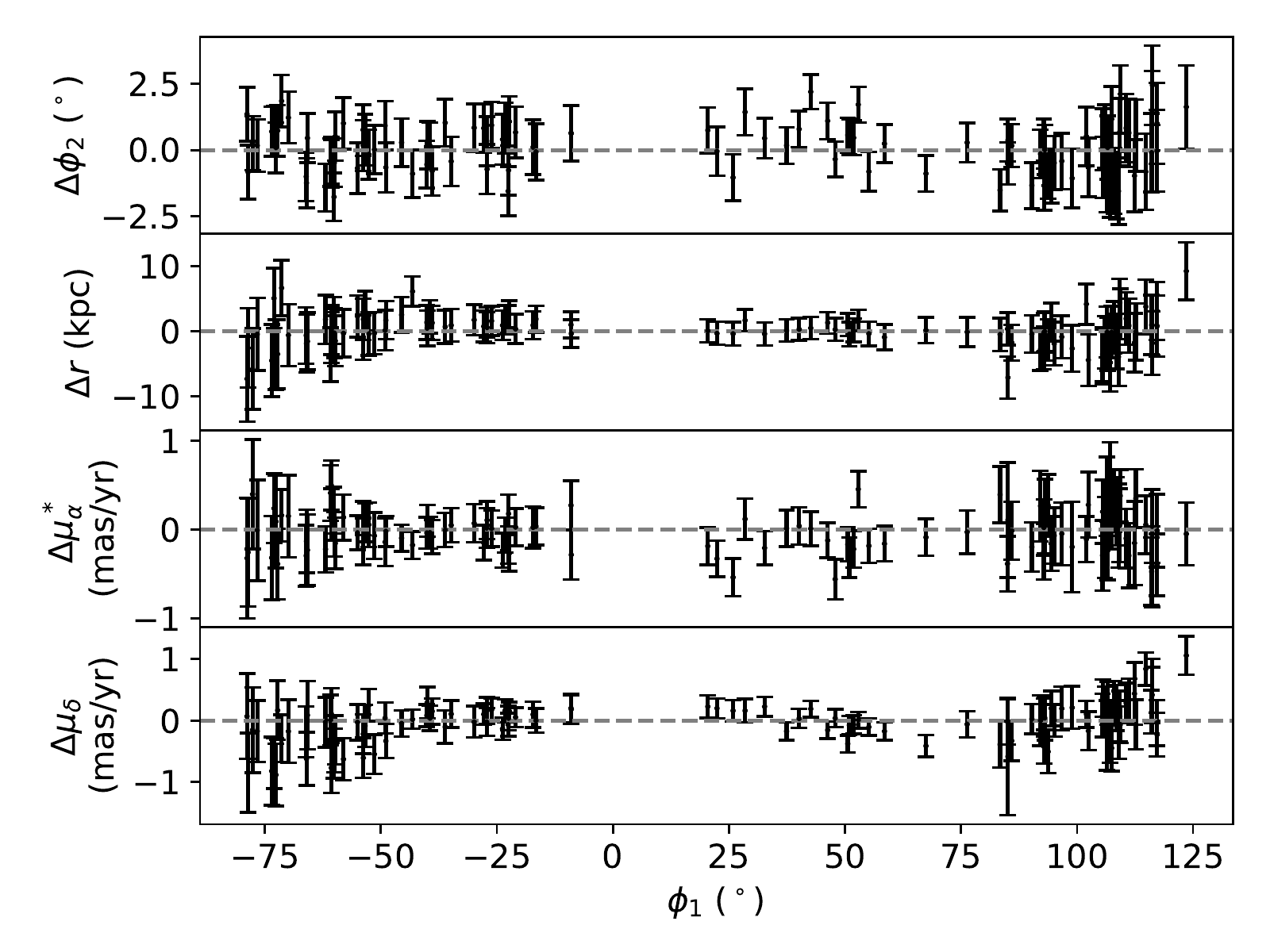}
\caption{Residuals of best fit model shown in \protect\figref{fig:fixed_MW_bestfit}. In each panel we plot the model value minus the observed value and highlight the value of zero with a dashed-grey line. The error bars come from summing the model width and observed errors in quadrature. There are no significant residuals apart from an offset in $\Delta \mu_\delta$ between $50^\circ < \phi_1 < 75^\circ$.  } 
\label{fig:residuals}
\end{figure}

\Figref{fig:residuals} shows the residuals of the best-fit model from
\Figref{fig:fixed_MW_bestfit}. For each RR Lyrae and each observable,
we plot the model value minus the observed value. Overall, there are
almost no significant residuals showing that the best-fit model is a
good representation of the data. However, there is a small residual in
$\Delta \mu_\delta$ between $50^\circ < \phi_1 <
75^\circ$. Interestingly, this lines up with an underdensity seen in RR Lyrae \citep[see Fig. 15 in][]{koposov_etal_orphan}.

\begin{figure}
\centering
\includegraphics[width=0.49\textwidth]{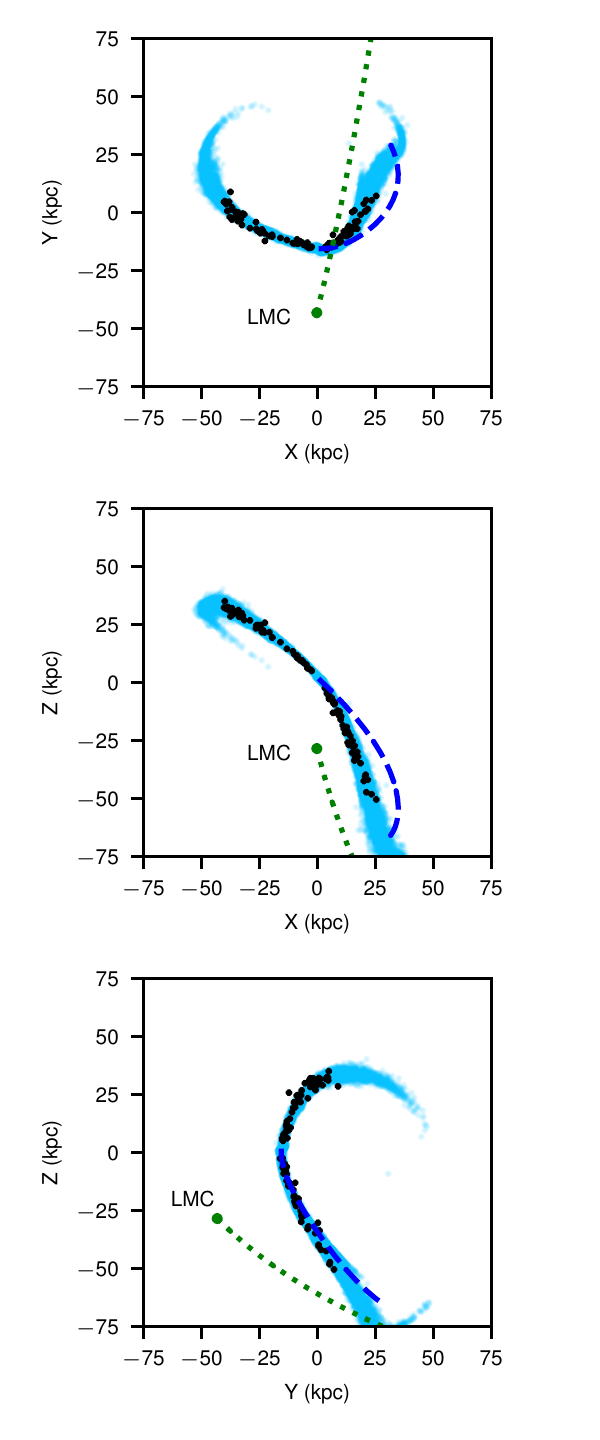}
\caption{Best fit model from \protect\figref{fig:fixed_MW_bestfit} shown in 3d in Galactocentric coordinates. From top to bottom, the panels show the XY, XZ, and YZ projections. In each panel, the light blue points show the model stream, the black points show the RR Lyrae positions, the dashed-blue lines show the orbit of the progenitor, and the dotted green line shows the orbit of the LMC. The LMC's nearby passage pulls the stream away from its orbit which leads to the offset seen in \protect\figref{fig:orphan_data}.} 
\label{fig:fixed_MW_bestfit_3d}
\end{figure}

As an alternative way to visualize the effect of the LMC, we show the
3D projections of the OS data (black filled circles), our
best-fit OS model (blue points), the corresponding progenitor's orbit
(dashed blue line) and the LMC's orbit (dotted green line) in
\Figref{fig:fixed_MW_bestfit_3d}. The present day position of the LMC
is marked with large filled green circle. The three panels show the
projections in Galactocentric Cartesian coordinates. As the Figure
demonstrates, the model stream lines up with the progenitor's orbit in
the Galactic North. However, in the South, the OS appears
to be pulled away from its orbit towards the orbit of the LMC. Of
course, these projections only show the present day location of the
stream. A movie showing the disruption of the OS in a static, spherical Milky Way halo in the
presence of the LMC can be found at
\href{https://youtu.be/sBKpwQR7JJQ}{https://youtu.be/sBKpwQR7JJQ}.

\begin{figure*}
\centering
\includegraphics[width=0.95\textwidth]{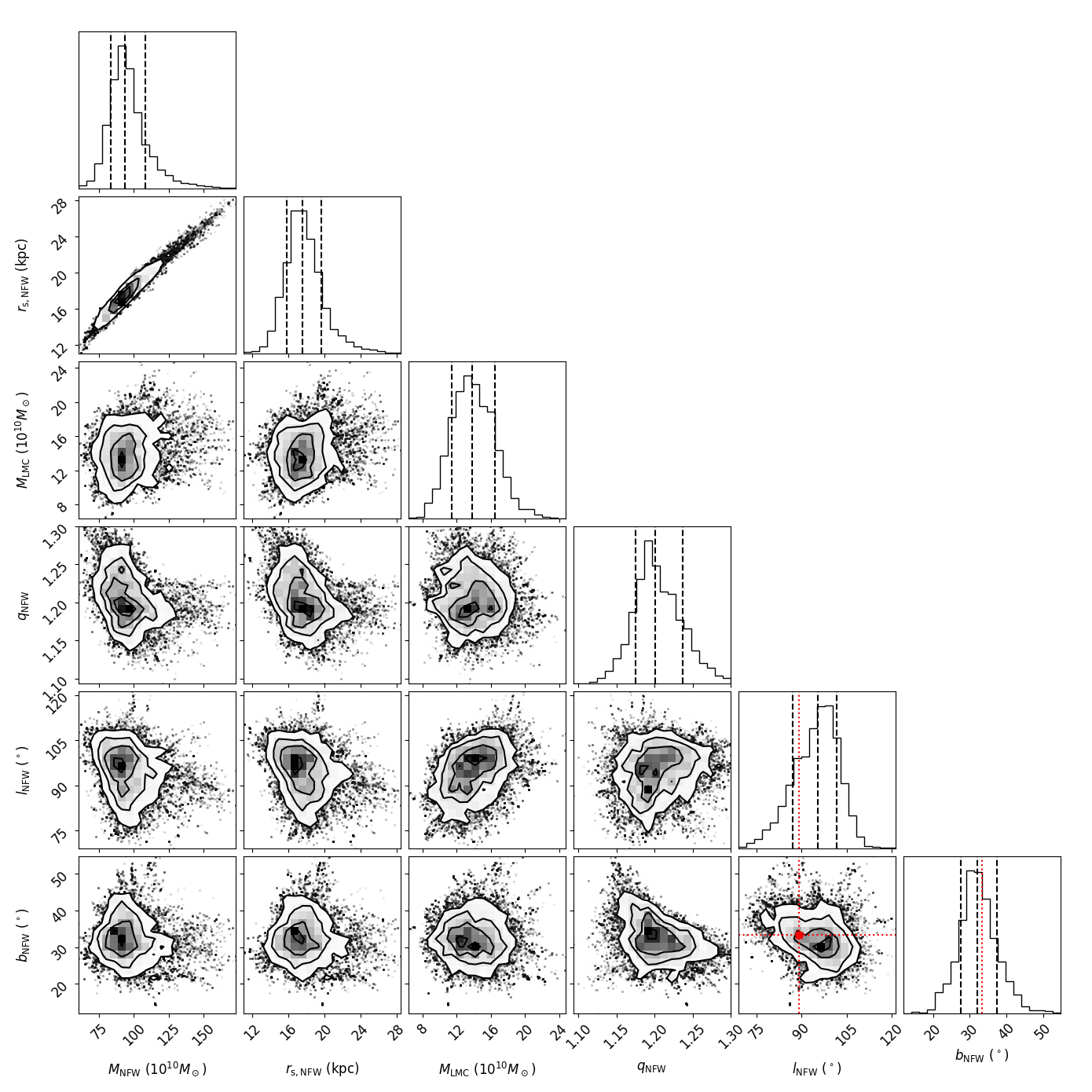}
\caption{Posteriors in fixed, prolate Milky Way potential with LMC. In each of the diagonal posteriors, we show the median with the $\pm1 \sigma$ range.  The red point in the bottom right figure shows the direction corresponding to the LMC's present day location from the galactic center. Interestingly, the flattening is closely aligned with the direction towards the LMC suggesting that it may be related to the LMC.  } 
\label{fig:posteriors_prolate_fixed}
\end{figure*}

Figure \ref{fig:posteriors_prolate_fixed} show the posterior distributions of
the Milky Way and LMC parameters for our best-fit halo, namely a reflexive, prolate Milky Way
halo. The distributions of individual parameter values
look single-peaked and well-behaved. There is also a familiar
degeneracy between the scale-radius and the DM halo mass of the Milky
Way. This is likely because the OS is in the outskirts of
the Milky Way and thus it is not sensitive to the inner profile of the
halo. The fits favor a Milky Way mass of $9.4\times10^{11} \,\rm{M_\odot}$ 
although the posteriors are quite broad at the $2\sigma$ level. Most interestingly, we get a tight constraint on the
LMC mass of $1.38^{+0.24}_{-0.27}\times10^{11} \,\rm{M_\odot}$. Thus, for the first time we have a
robust measurement of the LMC mass from its effect on a Milky Way
stream.

The posteriors of all of our fits are given in \Tabref{tab:posteriors}. Curiously,
in the case of flattened haloes, the fits prefer a significantly aspherical halo:
in the prolate case, a flattening of $q_{\rm NFW}
\sim 1.20$ is inferred, while in the oblate case, the model converges
to $q_{\rm NFW} \sim 0.87$. In addition, the fits also prefer a particular orientation.  In the prolate case,
the halo is stretched towards roughly the present-day location of the
LMC. In the oblate case, the flattened axis of the halo is within
$\sim 20^\circ$ of the orbital plane of the LMC. Interestingly, the
orientation in the oblate case is consistent with the results
\cite{law_majewski_2010} who fit the Sagittarius stream and found an
almost axisymmetric, oblate halo flattened close to the LMC's orbital
plane. We will discuss the significance of these results more in
\Secref{sec:discussion}. The maximum-likelihood values for each of these setups are given in \protect\Tabref{tab:mlhood}.

\subsection{Orbit of Orphan and the LMC}

We find that the orbit of the Orphan stream progenitor changes significantly before and after the LMC's infall, and that it depends on how we treat the Milky Way halo. For example, in the case of a fixed, prolate halo, the orbit has a pericenter of $\sim 23$ kpc and an apocenter of $\sim 79$ kpc before the LMC's infall. Just after infall, this pericenter is reduced to $\sim 16$ kpc. In a prolate halo with a reflexive Milky Way, the pericenter and apocenter before the LMC's infall are $\sim 23$ kpc and $\sim$ 71 kpc respectively while the post-infall pericenter is $\sim 16$ kpc. In contrast, for a spherical, reflexive halo, the pre-infall pericenter and apocenter are $\sim 28$ kpc and $\sim 59$ kpc respectively, while the post-infall pericenter is again $\sim 16$ kpc. Thus we see that the Orphan progenitor was on a less eccentric orbit before interacting with the LMC. However, we note that since the effect of the LMC varies strongly along the stream (see \figref{fig:fixed_MW_bestfit}), the change in the orbit of each star will likewise vary along the stream. For example, stars in the South which have been most affected by the LMC have been pulled out beyond 95 kpc while stars in the North have been affected much less.

We have also checked whether the LMC has additional pericentric passages over the past 5 Gyr. In the prolate, reflexive halo, the median time at which the LMC reaches its apocenter is 3.6 Gyr ago but none of the chains have an LMC which has an additional pericenter in the past. In the spherical, reflexive halo, the median apocenter is 3.6 Gyr with 1.6\% of our chains having an additional pericenter close to 5 Gyr ago. Thus, in the vast majority of cases, the LMC only has a strong effect on Orphan in the recent past. In the few cases that the LMC does have an additional interaction, this will have little effect on the present day properties of Orphan in our model since only a small fraction of the stars in Orphan would have stripped by those early times. In addition, since the stream grows in time, these stars would be far from the progenitor and thus outside the observed region.

\subsection{Mass constraint on the Milky Way}

\begin{figure}
\centering
\includegraphics[width=0.45\textwidth]{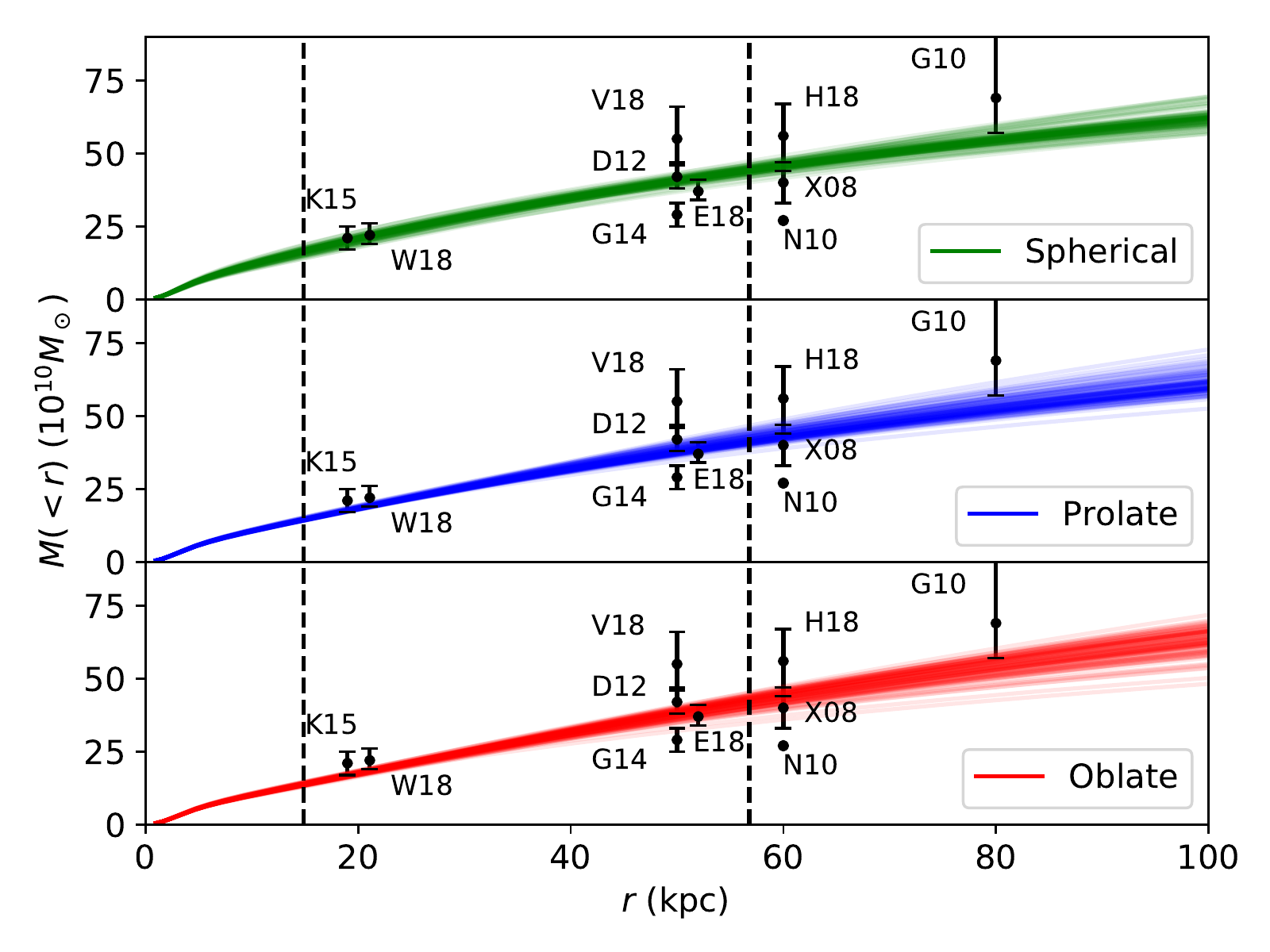}
\caption{Mass constraint for total Milky Way mass as a function of radius. From top to bottom, we show samples from our MCMC chains in a spherical, prolate, and oblate halo which can respond to the LMC. In each panel, the black points with error bars show measurements from \protect\cite{deason_etal_2012} (D12), \protect\cite{eadie2018} (E18) \protect\cite{gnedin_etal_2010} (G10), \protect\cite{mLCS} (G14), \protect\cite{hendel_etal_2018} (H18), \protect\cite{Kupper2015} (K15), \protect\cite{newberg_etal_2010} (N10), \protect\cite{vasiliev_2018} (V18),  \protect\cite{watkins_etal_2018} (W18), and \protect\cite{Xue2008} (X08). Note that N10 did not provide an error bar on their result and we have shifted E18 from 50 kpc for clarity. The dashed-black vertical lines show the radial extent of the Orphan RR Lyrae sample from \protect\cite{koposov_etal_orphan} fit in this work. } 
\label{fig:mw_mass}
\end{figure}

Finally, we compare our constraint on the Milky Way mass profile with
existing results in \Figref{fig:mw_mass}. Despite the broad posteriors in \Figref{fig:posteriors_prolate_fixed}, the
constraint on the Milky Way mass as a function of radius is remarkably
tight. Our results also agree with existing results in the literature,
although with a tendency to prefer lower values of the total mass.
Note that when computing the mass enclosed in \Figref{fig:mw_mass}, 
we have ignored the flattening of the halo. For reference, we also give the
mass enclosed within 50 kpc in \Tabref{tab:posteriors}. As expected from \Figref{fig:mw_mass},
this mass is remarkably well constrained. This is due to the wealth of RR Lyrae uncovered
by \cite{koposov_etal_orphan} and the precise multi-dimensional data from \textit{Gaia} DR2.

\section{Discussion} \label{sec:discussion}

\subsection{Limitations of this analysis}

In this work we have shown that the observed phase-space properties of
the OS can be well-modelled if we include the effect of the
LMC. We have attempted to fit Orphan without the LMC and the
result is rather spectacularly poor (see \figref{fig:model_comparison})
due to the misalignment of the stream with its proper motions. While
we only considered one form for the potential, we argue that in
general, this offset cannot be reproduced in a static potential. In
this context, we note that \citep{sanders_binney_2013} studied streams
in action-angle space and showed that there is a misalignment between
streams and orbits in action-angle space. However, they found that in
their chosen Milky Way potential, the OS should have the
smallest misalignment. We can also compare the size of the velocity
component perpendicular to the stream \citep[$\sim 60$ km/s from][]{koposov_etal_orphan}
to the expected velocity dispersion of the progenitor at its tidal
radius. Assuming a progenitor mass of $10^7 \,\rm{M_\odot}$ and a pericenter
of 15 kpc, this would correspond to a tidal radius of 493 pc and thus
a velocity dispersion of $\sim 5$ km/s in \texttt{MWPotential2014}
from \cite{galpy}. Since this is significantly smaller than the
measured velocity offset, we do not think it is possible to produce
such an offset in any static potential. Note that our assumed progenitor mass
is based on the total stellar mass in the OS estimated in \cite{koposov_etal_orphan}. 
Although the OS progenitor was much more massive than this originally, by the time the stellar stream formed, almost all of the dark matter would have already been stripped. 

\begin{figure}
\centering
\includegraphics[width=0.45\textwidth]{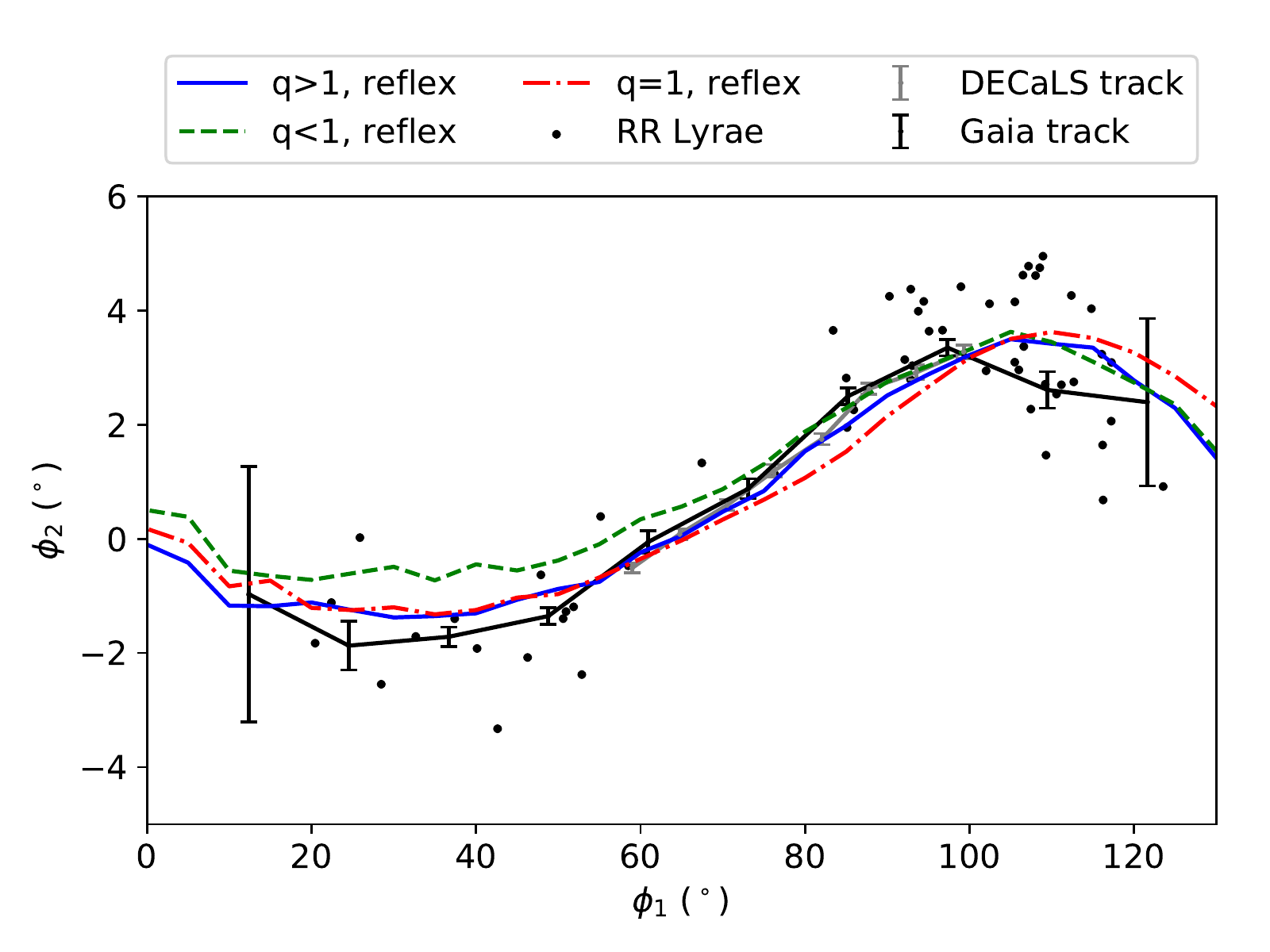}
\caption{Comparison of RR Lyrae and stream track from \protect\cite{koposov_etal_orphan} with our best-fit models. }
\label{fig:stream_track}
\end{figure}

In the analysis in Sections \ref{sec:mw_fit}, \ref{sec:lmc_fits} we have fit the RR
Lyrae sample from \cite{koposov_etal_orphan}. However, \cite{koposov_etal_orphan}
also measured the track using Red Giant Branch (RGB) stars from
\textit{Gaia} DR2 and RGB/Main Sequence Turn-Off stars in DECaLS. We
show these tracks with our best-fit streams in
\Figref{fig:stream_track}. Interestingly, we see that in the regions
where the tracks and RR Lyrae positions appear to slightly disagree
($\phi_1 \sim 100^\circ$), our best-fit models appear to follow the
tracks in these other tracers. Since both the prolate and oblate haloes are a better fit to
the RR Lyrae in this range, this could suggest that fits to the stream
tracks (instead of the individual RR Lyrae positions) might prefer a
more spherical halo.

\subsection{Interpretation of the LMC mass}

In this work we have shown that including an LMC with a substantial
mass is essential for understanding the behavior of the OS. The models used in this work have all treated the LMC mass as
being fixed in time. In reality, as the LMC falls onto the Milky Way,
the dark matter in its outskirts will be tidally stripped, changing
the mass profile and shape of the LMC. Note that this stripped
material will still exert a force on the Milky Way and the OS so accounting for the stripping is
not as simple as just removing the mass. Accordingly, we surmise that
this work constrains an effective mass for the LMC which captures its
effect on the OS. Since the LMC is believed to be on its
first approach \citep[e.g.][]{besla07,lmc_pms}, this mass should not
be too different from the peak mass of the LMC although future work is
needed to better understand their relation.

\subsection{Interpretation of Milky Way halo shape} \label{sec:mw_halo_shape}

The best-fit models to the OS in \Secref{sec:lmc_fits} require that the Milky Way halo must be
substantially aspherical. We considered
axisymmetric haloes with a flattening in an arbitrary direction. For
prolate haloes, we found a halo with an elongation that is roughly
aligned with the present day position of the LMC. For oblate haloes,
we found a halo with a flattening roughly aligned with the orbital
plane of the LMC. The oblate halo shape is especially interesting
given its similarity with \cite{law_majewski_2010} who studied the
Sagittarius stream and found a very similar orientation. Thus, the
only two streams that probe the outskirts of the Milky Way both point
to a similarly flattened halo in the outskirts. If this is the case,
it would have important implications for the plane of satellites since
this plane can be long-lived if it is aligned with the short or long
axis of the halo \citep{bowden_etal_2013}. We note that although
\cite{law_majewski_2010} considered triaxial haloes, they forced the
halo to have one flattening in the Galactic $z$ direction so we cannot
directly compare their result with our prolate halo.

\begin{figure*}
\centering 
\includegraphics[width=0.95\textwidth]{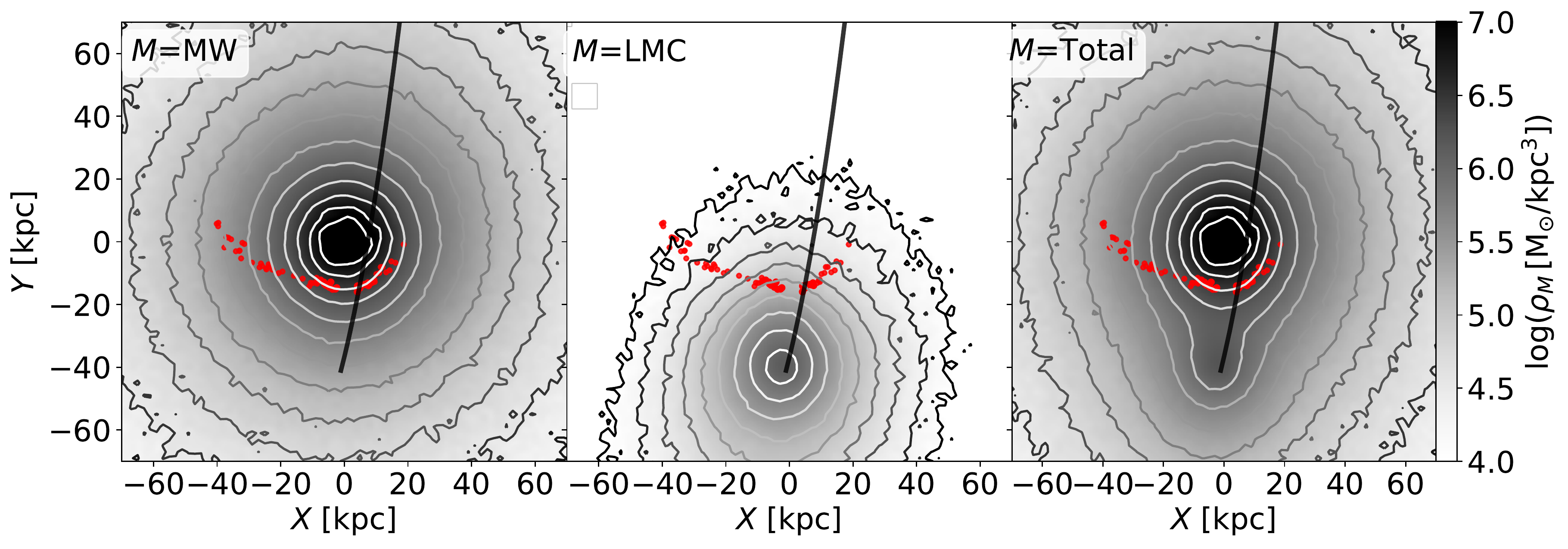}\\
\includegraphics[width=0.95\textwidth]{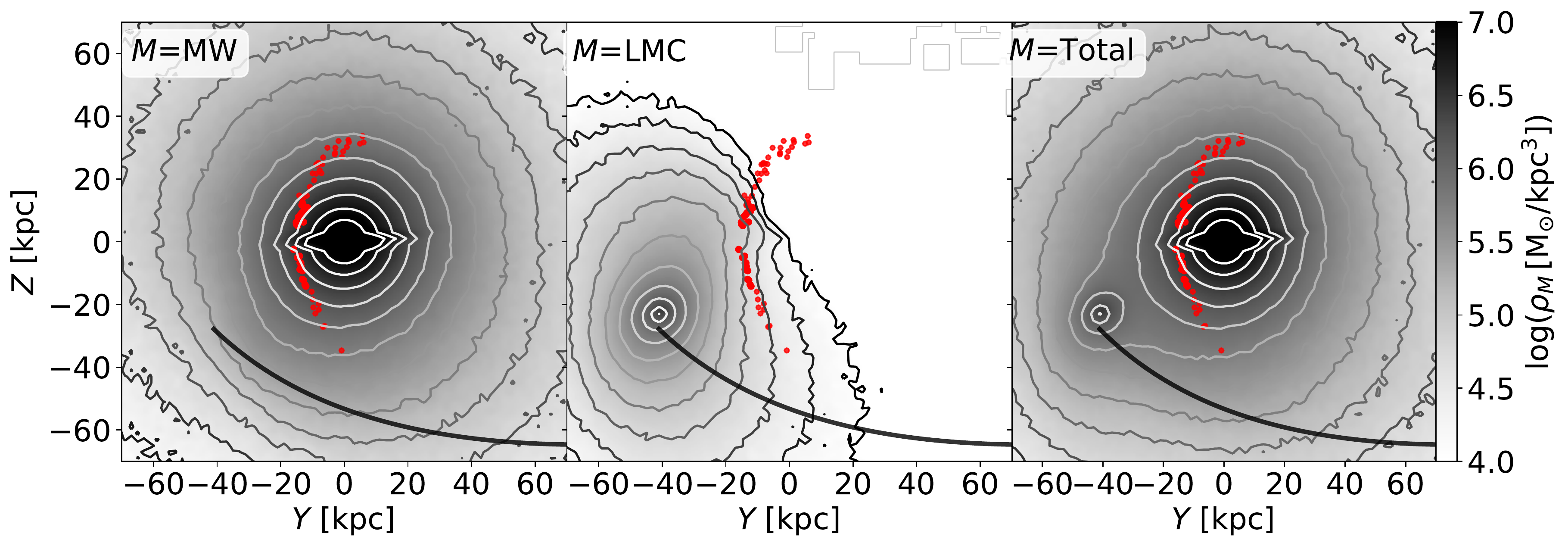}
\caption{Density of Milky Way and LMC in $N$-body simulation from \protect\cite{laporte2018}. The top three panels show an XY projection of a 20 kpc thick slab centered on $Z=0$ in Galactocentric coordinates. The bottom three panels show YZ projections show a 20 kpc thick slab centered on $X=0$. The left, middle, and right panels show projections of the Milky Way, LMC, and the combined Milky Way-LMC densities respectively. The orbit of the LMC is shown as a black line and the RR Lyrae from \protect\cite{koposov_etal_orphan} are shown as red points. Although the LMC is heavily distorted by the Milky Way (middle panels), the Milky Way itself remains remarkably spherical on average (left panels).  This is in contrast to our best-fit prolate and oblate models which have significant flattenings of $q=1.20$ and $q=0.87$ in the potential, respectively (see \protect\tabref{tab:posteriors}).}
\label{fig:nbody_sims}
\end{figure*}

In order to better understand the effect of the LMC on the Milky Way halo, 
we show projections of an $N$-body simulation of the LMC accreting onto the
Milky Way in \Figref{fig:nbody_sims}. These simulations are taken from
\cite{laporte2018} who simulate a $2.5\times 10^{11} \,\rm{M_\odot}$ LMC on its first 
infall onto the Milky Way. The LMC in this model ends up within $2\sigma$
of its observed phase-space position. The top (bottom) panels show the XY (YZ) projection
of a 20 kpc slab centered on Z=0 (X=0) from the snapshot which matches the LMC at the present day. 
Interestingly, the Milky Way appears to be mostly spherical (left panels) while the LMC has
been significantly tidally disrupted by the Milky Way (middle panel). The orbit of the LMC
and the RR Lyrae from \cite{koposov_etal_orphan} are also shown for reference. 

While we have shown that the OS requires a substantially
aspherical halo shape for the Milky Way, we must be careful in
interpreting the meaning of this. One possibility is that this
reflects the long term shape of the Milky Way halo. Indeed, simulations
of Milky Way-like galaxies have shown that haloes can have axis ratios of 0.6-0.8 in density
even when baryonic processes are included \citep[e.g.][]{zhu_etal_2016}. Such a flattening is consistent
with our oblate haloes which have a flattening of $q\sim0.9$ in the potential. However, in this case, the
Milky Way disk would need to be aligned with the short or long axis of
the halo to be stable \citep{debattista_etal_2013}. This would be an
issue for our prolate halo solution which is substantially misaligned
with the disk. However, since Orphan is only probing the halo beyond
$\sim 10$ kpc, the halo could have a different shape in the inner
regions and the shape we find here in the outskirts. Thus, the disk
could still be made stable in this scenario.

A second possibility is that the shape we find here reflects the
response of Milky Way's halo to the infall of the LMC
\citep{weinberg89}. This effect was investigated in
\cite{vesperini_weinberg_2000} who showed that the infall of
satellites can induce substantial density features in the host
halo. Similarly, \cite{gomez_etal_2016} investigated accretion events
in cosmological zoom-in simulations and found that the density of the
host halo can be substantially perturbed which would lead to torques
on structures within the host halo larger than those expected from the
perturber itself. Despite the high relative velocity of the LMC, the
same mechanism is still able to operate on the the Milky Way, notably warping
the disk \citep{laporte2018}. This interpretation is further supported
by the fact that the halo shapes we infer are aligned with either the
present day position of the LMC (prolate halo) or with the orbital
plane of the LMC (oblate halo). However, as we can see from \Figref{fig:nbody_sims}, 
this is not expected to generate an extremely flattened Milky Way. 

Finally, it is possible that the halo shape we infer is not due to the
Milky Way but is rather capturing an effect from the LMC
which is not correctly modelled by treating the LMC as a non-deforming
Hernquist sphere. As with the previous explanation, this is supported
by the fact that the halo shapes are related to either the LMC's
present day position or orbital plane. One effect we are neglecting is the tidal 
disruption of the LMC by the Milky Way which
will dramatically stretch the dark matter distribution in the
outskirts of the LMC (e.g. middle panels of \figref{fig:nbody_sims}). We are also ignoring the SMC which could have an
additional affect on Orphan. However, as we argued above, since the
SMC has already been substantially disrupted by the LMC, its present-day mass
should be quite small.

\subsection{The impact of the LMC on other structures in the Milky Way}

For the first time, we have measured the significant impact of the LMC
on a structure in the Milky Way. The LMC has three main effects on
structures orbiting our Galaxy. First, structures which pass near the
LMC will feel a large force which will directly change their
dynamics. Second, the LMC will induce a reflex motion in the Milky Way
itself \citep{gomez_etal_2015} which can affect structures in our
Galaxy. Third, the LMC can deform the Milky Way halo, resulting in
additional tidal forces \citep[e.g.][]{weinberg98,weinberg06}.

These effects have wide ranging implications for any technique which
attempts to precisely measure the properties of the Milky Way by
studying tracers orbiting our Galaxy, e.g. equilibrium modeling of GCs
\citep[e.g.][]{watkins_etal_2010,vasiliev_2018}, the stellar halo
\citep[][]{Xue2008,deason_etal_2012}, or hypervelocity stars
\citep{gnedin_hvs}. All of these techniques should be revisited in
light of the increased LMC mass to understand how the results may be
biased by ignoring the presence of our largest - and most massive -
satellite.

Along these lines, \cite{laporte2018} have studied the effect of the
LMC on its first infall \citep{besla07} on the Milky Way's disk using
live high-resolution N-body simulations. They found that the response
of the halo can result in strong overdensities of up to $50\%$ within
$\sim 40 \rm{kpc}$ penetrating into the inner regions of the Galaxy,
resulting in torques warping the disk with similar shape and line of
nodes as the observed HI warp \citep{levine06}, in agreement with
earlier expectations from linear perturbation theory
\citep[e.g.][]{weinberg98,weinberg06}. The impact of the Milky Way's dark
matter halo response to the LMC on tracer kinematics in the halo are
explored further in \cite{garavito_camargo_etal_2019}. 

Given the large effect seen in this work, we expect that many other
streams will exhibit similar effects from the LMC. The streams in the
Southern Galactic hemisphere are natural candidates since they have
had closer passages with the LMC. Tucana III is an especially good
candidate given its close passage with the LMC
\citep{tuciii_modelling,simon_2018,fritz_pms_2018}, as are the 11
streams recently discovered in the Dark Energy Survey
\citep{des_streams}. Since these streams will each have a different
closest approach to the LMC, they can be used to measure the LMC's
radial density profile and shape. Such modelling would not be simple
as it would need to account for the time-dependent LMC shape due to
its disruption by the Milky Way.

The substantial LMC mass found in this work suggests that the LMC has brought a large number of satellite galaxies into the Milky Way \citep{sales2011}. Indeed, the LMC group infall was proposed by \cite{jethwa_lmc_sats} to explain the abundance of satellites found in the Dark Energy Survey. Recently, \cite{kallivayalil_lmc_sats} confirmed this prediction and found 4 ultra-faint dwarfs which were likely accreted with the LMC, in addition to the SMC which was previously argued to have fallen in with the LMC \citep[e.g.][]{lmc_pms}. Given that \cite{jethwa_lmc_sats} predicted the LMC could have brought up to $\sim 70$ satellites, we expect that many more satellites will be found in the coming years. The phase-space positions of this large number of satellites will provide another avenue for measuring the LMC mass.

\begin{figure}
\centering
\includegraphics[width=0.45\textwidth]{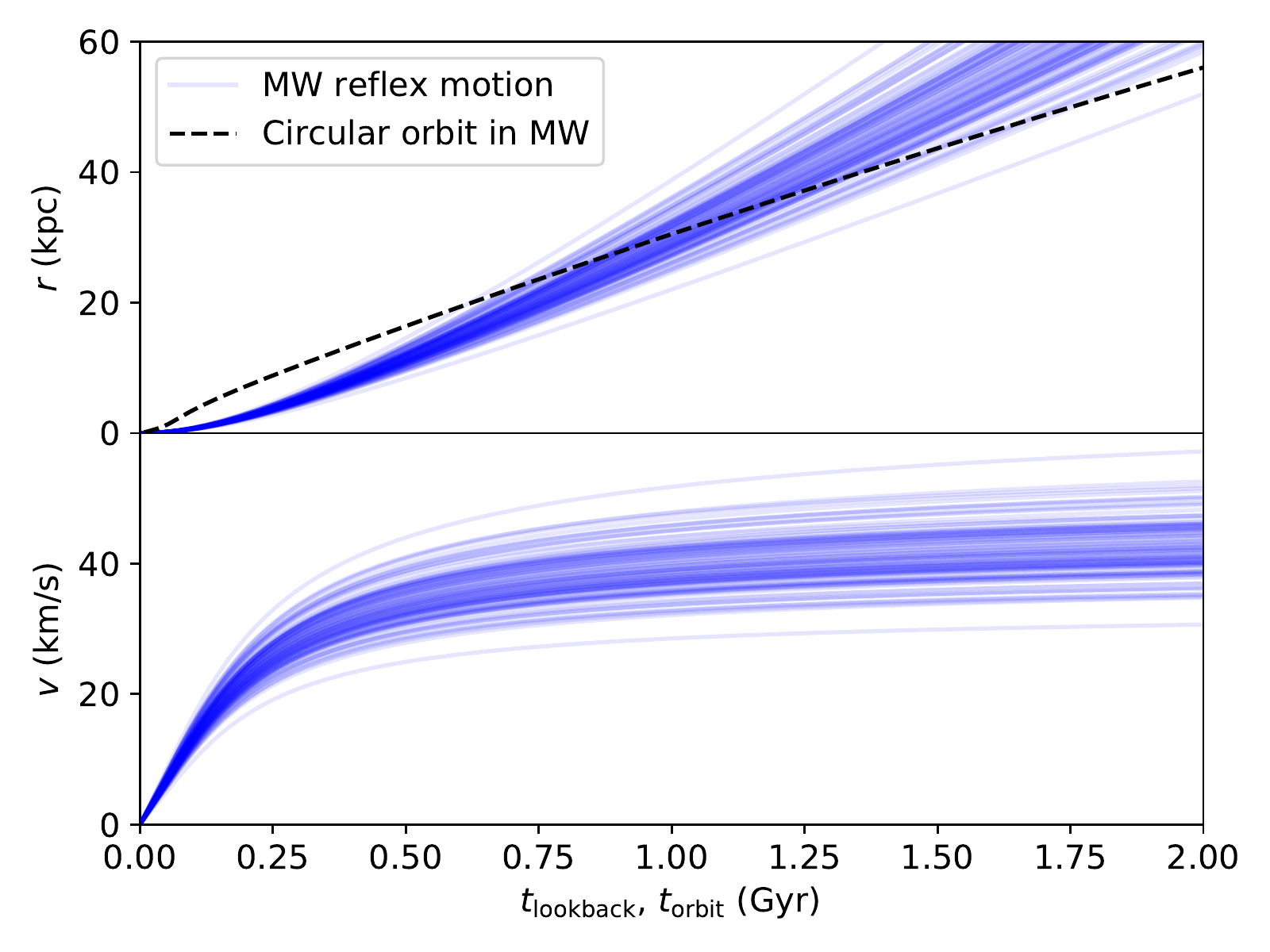}
\caption{Reflex motion of the Milky Way compared to orbital timescales in the Milky Way. The blue lines show the reflex motion of the Milky Way from 100 samples of the chains in the prolate halo fits. In the top (bottom) panel, these blue lines show the distance (speed) of the Milky Way moves relative to its present day position (velocity). The dashed-black line shows the orbital period versus radius for circular orbits in the Milky Way. Stars on orbits to the left of the blue lines can respond adiabatically to the LMC's infall and thus should move with the Milky Way's reflex velocity. However, stars on orbits to the right of the blue curves will not have time to respond to the Milky Way's motion and will remain on their original orbits. } 
\label{fig:reflex_motion}
\end{figure}

Finally, we can use the results of this work in order to assess the
importance of the reflex motion of the Milky Way. In
\Figref{fig:reflex_motion} we compare the expected reflex motion over
the past 2 Gyr with the orbital timescales in the Milky Way. The blue
lines show the reflex motion of the Milky Way in 100 samples of the
MCMC chains used to fit Orphan in a prolate Milky Way halo. These
should be compared with the dashed-black curve which shows the orbital
period versus radius for circular orbits in the Milky Way in
\texttt{MWPotential2014} from \cite{galpy}. Stars for which the
orbital timescale is much smaller than the timescale over which the
Milky Way moves by a distance equal to the orbital radius (i.e. those
to the left of the blue curves) will respond adiabatically to the
LMC's passage and will move with the reflex velocity of the Milky
Way. However, stars on orbits for which the orbital timescale is
longer than the timescale over which the Milky Way moves by their
orbital radius (i.e. those to the right of the blue curves) will
continue on their original orbits and should consequently have a velocity
relative to the Milky Way itself. We thus expect that there is a
region within the Milky Way ($r<\sim 30$ kpc) which will respond
adiabatically during the LMC's infall. However, beyond this radius,
the stars will not have had time to adjust to the Milky Way's velocity
after the LMC's infall and should have a velocity ($\sim 40$ km/s)
relative to interior part of the Milky Way. Since the majority of the
Milky Way's reflex motion is downwards, we predict that the outer
regions of the Milky Way should have an upwards velocity relative to
the inner regions. This effect should be visible in the outskirts of
the Galactic stellar halo.

\section{Conclusions} \label{sec:conclusions}

Motivated by the new observational constraints recently presented by
\cite{koposov_etal_orphan}, this Paper presents the results of a
comprehensive modelling of the Orphan stream. Our analysis relies on the
rapid and accurate model stream production implemented using the mLCS
technique \citep[][]{mLCS} and considers a wide range of Galactic
Dark Matter halo configurations. Taking advantage of the efficiency
and the flexibility of our stream models, we are also able to include
the presence of the Large Magellanic Cloud as well as the reflex
motion of the Milky Way in response to the dwarf's infall. For the
first time, we have unveiled an unambiguous perturbation exerted by
the Large Magellanic Cloud on a stellar stream within the Milky
Way. In our reconstruction of the encounter, as the LMC was falling
into the Galaxy, it passed near the Orphan stream and pulled the stream off
its original path. The resulting deflection is so large that it can be
seen in the data itself as a misalignment of the Orphan stream track
and the direction of its proper motions (see
\figref{fig:orphan_data}). Detecting this effect was only possible
with the exquisite data quality in \textit{Gaia} DR2 which offers an
all-sky view of the RR Lyrae across the entire Milky Way, including
their proper motions.

Without the presence of the LMC, it is only possible to reproduce the
phase-space track of the stream in the North, i.e. for
$\phi_1>50^{\circ}$. The Southern Galactic portion of such a model is
an extremely poor fit to the data (see \figref{fig:mw_fit}). However,
taking the same model and including an LMC with a modest mass results
in a significant deflection of the Southern portion of the
stream. Furthermore, if the LMC mass is increased to $\sim10^{11}
\,\rm{M_\odot}$, this deflection grows and the resulting stream is a good
match to the Orphan data over the entire range of along-stream
coordinate $\phi_1$. Interestingly, this exercise shows that while the
LMC's main effect is on the Southern portion of Orphan, there is also
a noticeable deflection of the Northern OS debris suggesting that the 
LMC's influence will be felt across the Galaxy.

In order to measure the LMC mass, we fit the entire range of the
Orphan stream including the LMC mass as a parameter in our fits. These
fits are performed with an axisymmetric Milky Way halo which can be
flattened in any direction. The fits require LMC masses between
$1-1.5\times10^{11} \,\rm{M_\odot}$ depending both on whether we have a
spherical, oblate, or prolate halo and on whether we allow the Milky
Way halo to respond to the LMC's infall. Curiously, the fits in an
oblate and prolate halo prefer halo shapes which appear to be aligned
with either the LMC's orbital plane or the LMC's present day location
respectively. In these cases, we suspect that the Milky Way halo may
be compensating for some aspect of the LMC's effect which we have not
correctly modelled, e.g. the tidal disruption of the LMC by the Milky
Way. This may suggest that the LMC is more massive than our results imply, 
although we note that fits in a spherical halo can
also match the Orphan stream.

The implications of our measurement of the LMC mass are far reaching. 
Any technique which has assumed that the Milky Way is in equilibrium 
should be revisited to determine how the LMC will bias its results. As one
concrete example of such an effect, we predict that the outskirts of the 
Milky Way ($r>\sim 30$ kpc) will have a bulk upwards velocity ($\sim 40$ km/s) relative 
to the Sun due to the reflex response of our Galaxy. If the results of this work
are correct, future efforts with \textit{Gaia} DR2 should reveal a significantly
perturbed Milky Way.  

We note that in the final stages of preparing this manuscript, \cite{fardal2018} released their analysis of the Orphan stream. Their work focused on the Northern portion of the Orphan stream and showed that there was a misalignment between the stream track and its velocity vector. 

\section*{Acknowledgements}

We thank the streams group at Cambridge and Eugene Vasiliev for helpful comments which improved the clarity of this work. DE thanks Michelle Collins and Shu Kim for insightful discussions. 
We thank the anonymous referee for their thoughtful comments which have improved the clarity of this work.
This work was started at the Aspen Center for Physics, which is
supported by National Science Foundation grant PHY-1607611. We thank
Michael for an illuminating discussion. The research leading to these
results has received funding from the European Research Council under
the European Union's Seventh Framework Programme (FP/2007-2013) / ERC
Grant Agreement n. 308024. SK is partially supported by NSF grant
AST-1813881. NK is supported by NSF CAREER award 1455260.

This work presents results from the European Space Agency (ESA) space
mission Gaia. Gaia data are being processed by the Gaia Data
Processing and Analysis Consortium (DPAC). Funding for the DPAC is
provided by national institutions, in particular the institutions
participating in the Gaia MultiLateral Agreement (MLA). The Gaia
mission website is https://www.cosmos.esa.int/gaia. The Gaia archive
website is https://archives.esac.esa.int/gaia.

This research made use of \textsc{ipython} \citep{IPython}, python packages \textsc{numpy} \citep{numpy}, \textsc{matplotlib} \citep{matplotlib}, and \textsc{scipy} \citep{scipy}. This research also made use of Astropy,\footnote{http://www.astropy.org} a community-developed core Python package for Astronomy \citep{astropy:2013, astropy:2018}.

\bibliographystyle{mn2e_long}
%\newpage
\bibliography{citations_orphan}

\appendix

\section{Best-fit parameters}

In this appendix, we give the best-fit (i..e maximum likelihood) parameters for each setup considered in \Tabref{tab:posteriors}.

\begin{table*}
\begin{centering}
\begin{tabular}{|c|c|c|c|c|c|c|}
\hline
Parameter  &  sph. MW$+$LMC    & obl. MW$+$LMC        & pro. MW$+$LMC     & sph. rMW$+$LMC  & obl. rMW$+$LMC  & pro. rMW$+$LMC \\
\hline Orphan & & & & & &  \\
\hline
$\mu_{\alpha,\, \rm prog}^*$ (mas/yr)  & -3.681  & -3.779 & -3.753 &-3.653 &-3.616 & -3.796  \\
$\mu_{\delta,\, \rm prog}$ (mas/yr) & 2.880 & 2.965 & 2.976 & 2.832 & 2.869 & 2.982 \\
$v_{r,\, \rm prog}$ (km/s) & 107.573 & 111.877 & 105.708 & 97.055 & 111.716 &  97.590 \\
$d_{\rm prog}$ (kpc) & 18.764 & 17.889 & 18.057 &18.350 & 17.833 & 17.846 \\
$\phi_{\rm 2,\,prog}$ ($^\circ$)  & -0.441 & -0.133 & -0.657 & -0.402 & -0.857 & -0.867 \\
\hline Milky Way  & & & & & &  \\
\hline $M_{\rm NFW}$ ($10^{10} M_\odot$)  & 123.643 & 125.839 & 108.360 &79.160 & 96.471 & 82.225 \\
$ r_s$ (kpc)  &17.707 & 21.244 & 18.902 & 12.807 & 19.550 & 15.013 \\
$q_{\rm NFW}$  & $-$ & 0.898 & 1.227 & $-$ & 0.851 & 1.272 \\
$l_{\rm NFW}$ ($^\circ$) & $-$ & 13.379 & 89.548 & $-$ & -39.448 &  94.759 \\
$b_{\rm NFW}$  ($^\circ$) & $-$ & -3.736 & 29.556 & $-$ & 9.001 & 29.145 \\
\hline LMC  & & & & & & \\
\hline $M_{\rm LMC}$ ($10^{10} M_\odot$) & 7.173 & 10.464 & 8.391 &12.501 & 9.038 & 12.657  \\
$\mu_{\alpha,\, \rm LMC}^*$ (mas/yr)  & 1.906 & 1.883 & 1.910& 1.888 &1.909 & 1.930 \\
$\mu_{\delta,\, \rm LMC}$ (mas/yr) & 0.435 & 0.315 & 0.308 & 0.293 & 0.221 & 0.307 \\
$v_{r,\, \rm LMC}$ (km/s)  & 261.286 & 261.667 & 264.715 & 263.960 & 260.910 & 265.014 \\
$d_{\rm LMC}$ (kpc) & 48.009 & 50.537 & 50.999 & 47.893 & 49.241 &  52.184 \\ \hline   
\end{tabular}
\caption{Best-fit values in the 6 setups from \protect\tabref{tab:posteriors}. For ease in reading, we give these in the same format as \protect\tabref{tab:priors} where we described the priors.  }
\label{tab:mlhood}
\end{centering}
\end{table*}

\end{document}

%% file: authors.tex
\author[Erkal et al.]{
\parbox{\textwidth}{
\Large
D.~Erkal,$^{1}$\thanks{d.erkal@surrey.ac.uk}
V.~Belokurov,$^{2,3}$
C.~F.~P.~Laporte,$^{4}$
S.~E.~Koposov,$^{5,2}$
T.~S.~Li,$^{6,7}$
C.~J.~Grillmair,$^{8}$
N.~Kallivayalil,$^{9}$
A.~M.~Price-Whelan,$^{10}$
N.~W.~Evans,$^{2}$
K.~Hawkins,$^{11}$
D.~Hendel,$^{12}$
C.~Mateu,$^{13}$
J.~F.~Navarro,$^{14}$
A.~del~Pino,$^{15}$
C.~T.~Slater,$^{16}$
and S.~T.~Sohn$^{15}$
\begin{center} (The OATs: Orphan Aspen Treasury Collaboration) \end{center}
}
\vspace{0.4cm}
\\
\parbox{\textwidth}{
$^{1}$ Department of Physics, University of Surrey, Guildford GU2 7XH, UK\\
$^{2}$ Institute of Astronomy, University of Cambridge, Madingley Road, Cambridge CB3 0HA, UK\\
$^{3}$ Center for Computational Astrophysics, Flatiron Institute, 162 5th Avenue, New York, NY 10010, USA\\
$^{4}$ CITA National Fellow, Department of Physics and Astornomy, University of Victoria, 3800 Finnerty Road, Victoria, BC, V8P 5C2, Canada\\
$^{5}$ Carnegie Mellon University, 5000 Forbes Ave, Pittsburgh, PA, 15213, USA\\
$^{6}$ Fermi National Accelerator Laboratory, P.O.\ Box 500, Batavia, IL 60510, USA\\
$^{7}$ Kavli Institute for Cosmological Physics, University of Chicago, Chicago, IL 60637, USA\\
$^{8}$ IPAC, Mail Code 314-6, Caltech, 1200 E. California Blvd., Pasadena, CA 91125, USA\\
$^{9}$ Department of Astronomy, University of Virginia, 530 McCormick Road, Charlottesville, VA 22904, USA\\
$^{10}$ Department of Astrophysical Sciences, Princeton University, 4 Ivy Lane, Princeton, NJ 08544, USA\\
$^{11}$ Department of Astronomy, The University of Texas at Austin, 2515 Speedway Boulevard, Austin, TX 78712, USA\\
$^{12}$ Department of Astronomy and Astrophysics, University of Toronto, 50 St. George Street, Toronto, Ontario, M5S 3H4, Canada\\
$^{13}$ Departamento de Astronom\'{i}a, Instituto de F\'{i}sica, Universidad de la Rep\'{u}blica, Igu\'a 4225, CP 11400 Montevideo, Uruguay\\
$^{14}$ Senior CIfAR Fellow, Department of Physics and Astronomy, University of Victoria, 3800 Finnerty Road, Victoria, BC, V8P 5C2 Canada\\
$^{15}$ Space Telescope Science Institute, 3700 San Martin Drive, Baltimore 21218, USA\\
$^{16}$ Department of Astronomy, University of Washington, Box 351580, Seattle, WA 98195, USA\\
}
}

%% file: orphan.bbl
\begin{thebibliography}{127}
\expandafter\ifx\csname natexlab\endcsname\relax\def\natexlab#1{#1}\fi

\bibitem[{{Aaronson}(1983)}]{aaronson_1983}
{Aaronson} M., 1983, \apjl, 266, L11

\bibitem[{{Agnello} \& {Evans}(2012)}]{agnello_evans_2012}
{Agnello} A., {Evans} N.~W., 2012, \apjl, 754, L39

\bibitem[{{Amorisco}, {Martinez-Delgado} \& {Schedler}(2015){Amorisco},
  {Martinez-Delgado}, \& {Schedler}}]{Amorisco2015}
{Amorisco} N.~C., {Martinez-Delgado} D., {Schedler} J., 2015, arXiv e-prints,
  arXiv:1504.03697

\bibitem[{{Astropy Collaboration} {et~al}\mbox{.}(2013){Astropy Collaboration},
  {Robitaille}, {Tollerud}, {Greenfield}, {Droettboom}, {Bray}, {Aldcroft},
  {Davis}, {Ginsburg}, {Price-Whelan}, {Kerzendorf}, {Conley}, {Crighton},
  {Barbary}, {Muna}, {Ferguson}, {Grollier}, {Parikh}, {Nair}, {Unther},
  {Deil}, {Woillez}, {Conseil}, {Kramer}, {Turner}, {Singer}, {Fox}, {Weaver},
  {Zabalza}, {Edwards}, {Azalee Bostroem}, {Burke}, {Casey}, {Crawford},
  {Dencheva}, {Ely}, {Jenness}, {Labrie}, {Lim}, {Pierfederici}, {Pontzen},
  {Ptak}, {Refsdal}, {Servillat}, \& {Streicher}}]{astropy:2013}
{Astropy Collaboration} {et~al.}, 2013, \aap, 558, A33

\bibitem[{{Auger} {et~al}\mbox{.}(2010){Auger}, {Treu}, {Bolton}, {Gavazzi},
  {Koopmans}, {Marshall}, {Moustakas}, \& {Burles}}]{Auger2010}
{Auger} M.~W., {Treu} T., {Bolton} A.~S., {Gavazzi} R., {Koopmans} L.~V.~E.,
  {Marshall} P.~J., {Moustakas} L.~A., {Burles} S., 2010, \apj, 724, 511

\bibitem[{{Behroozi}, {Wechsler} \& {Conroy}(2013){Behroozi}, {Wechsler}, \&
  {Conroy}}]{behroozi_etal_2013}
{Behroozi} P.~S., {Wechsler} R.~H., {Conroy} C., 2013, \apj, 770, 57

\bibitem[{{Belokurov} {et~al}\mbox{.}(2017){Belokurov}, {Erkal}, {Deason},
  {Koposov}, {De Angeli}, {Evans}, {Fraternali}, \& {Mackey}}]{Belokurov2017}
{Belokurov} V., {Erkal} D., {Deason} A.~J., {Koposov} S.~E., {De Angeli} F.,
  {Evans} D.~W., {Fraternali} F., {Mackey} D., 2017, \mnras, 466, 4711

\bibitem[{{Belokurov} {et~al}\mbox{.}(2007){Belokurov}, {Evans}, {Irwin},
  {Lynden-Bell}, {Yanny}, {Vidrih}, {Gilmore}, {Seabroke}, {Zucker},
  {Wilkinson}, {Hewett}, {Bramich}, {Fellhauer}, {Newberg}, {Wyse}, {Beers},
  {Bell}, {Barentine}, {Brinkmann}, {Cole}, {Pan}, \& {York}}]{orphan_disc_v}
{Belokurov} V. {et~al.}, 2007, \apj, 658, 337

\bibitem[{{Belokurov} \& {Koposov}(2016)}]{BK2016}
{Belokurov} V., {Koposov} S.~E., 2016, \mnras, 456, 602

\bibitem[{{Belokurov} \& {Erkal}(2019)}]{CloudsArms}
{Belokurov} V.~A., {Erkal} D., 2019, \mnras, 482, L9

\bibitem[{{Besla} {et~al}\mbox{.}(2007){Besla}, {Kallivayalil}, {Hernquist},
  {Robertson}, {Cox}, {van der Marel}, \& {Alcock}}]{besla07}
{Besla} G., {Kallivayalil} N., {Hernquist} L., {Robertson} B., {Cox} T.~J.,
  {van der Marel} R.~P., {Alcock} C., 2007, \apj, 668, 949

\bibitem[{{Besla} {et~al}\mbox{.}(2012){Besla}, {Kallivayalil}, {Hernquist},
  {van der Marel}, {Cox}, \& {Kere{\v s}}}]{besla_etal_2012}
{Besla} G., {Kallivayalil} N., {Hernquist} L., {van der Marel} R.~P., {Cox}
  T.~J., {Kere{\v s}} D., 2012, \mnras, 421, 2109

\bibitem[{{Besla} {et~al}\mbox{.}(2016){Besla}, {Mart{\'{\i}}nez-Delgado}, {van
  der Marel}, {Beletsky}, {Seibert}, {Schlafly}, {Grebel}, \&
  {Neyer}}]{besla_etal_2016}
{Besla} G., {Mart{\'{\i}}nez-Delgado} D., {van der Marel} R.~P., {Beletsky} Y.,
  {Seibert} M., {Schlafly} E.~F., {Grebel} E.~K., {Neyer} F., 2016, \apj, 825,
  20

\bibitem[{{Bonaca} \& {Hogg}(2018)}]{Bonaca2018}
{Bonaca} A., {Hogg} D.~W., 2018, \apj, 867, 101

\bibitem[{{Bonaca} {et~al}\mbox{.}(2018){Bonaca}, {Hogg}, {Price-Whelan}, \&
  {Conroy}}]{Bonaca2018spur}
{Bonaca} A., {Hogg} D.~W., {Price-Whelan} A.~M., {Conroy} C., 2018, arXiv
  e-prints, arXiv:1811.03631

\bibitem[{{Bovy}(2015)}]{galpy}
{Bovy} J., 2015, \apjs, 216, 29

\bibitem[{{Bovy}, {Erkal} \& {Sanders}(2017){Bovy}, {Erkal}, \&
  {Sanders}}]{bovy2017}
{Bovy} J., {Erkal} D., {Sanders} J.~L., 2017, \mnras, 466, 628

\bibitem[{{Bowden}, {Belokurov} \& {Evans}(2015){Bowden}, {Belokurov}, \&
  {Evans}}]{Bowden2015}
{Bowden} A., {Belokurov} V., {Evans} N.~W., 2015, \mnras, 449, 1391

\bibitem[{{Bowden}, {Evans} \& {Belokurov}(2013){Bowden}, {Evans}, \&
  {Belokurov}}]{bowden_etal_2013}
{Bowden} A., {Evans} N.~W., {Belokurov} V., 2013, \mnras, 435, 928

\bibitem[{{Cappellari} {et~al}\mbox{.}(2006){Cappellari}, {Bacon}, {Bureau},
  {Damen}, {Davies}, {de Zeeuw}, {Emsellem}, {Falc{\'o}n-Barroso},
  {Krajnovi{\'c}}, {Kuntschner}, {McDermid}, {Peletier}, {Sarzi}, {van den
  Bosch}, \& {van de Ven}}]{Cappellari2006}
{Cappellari} M. {et~al.}, 2006, \mnras, 366, 1126

\bibitem[{{Carlberg}(2009)}]{Carlberg2009}
{Carlberg} R.~G., 2009, \apjl, 705, L223

\bibitem[{{Carlberg} \& {Grillmair}(2013)}]{carlberg_gd1}
{Carlberg} R.~G., {Grillmair} C.~J., 2013, \apj, 768, 171

\bibitem[{{Carlberg}, {Grillmair} \& {Hetherington}(2012){Carlberg},
  {Grillmair}, \& {Hetherington}}]{carlberg_pal5}
{Carlberg} R.~G., {Grillmair} C.~J., {Hetherington} N., 2012, \apj, 760, 75

\bibitem[{{Choi} {et~al}\mbox{.}(2018){Choi}, {Nidever}, {Olsen}, {Besla},
  {Blum}, {Zaritsky}, {Cioni}, {van der Marel}, {Bell}, {Johnson}, {Vivas},
  {Walker}, {de Boer}, {No{\"e}l}, {Monachesi}, {Gallart}, {Monelli},
  {Stringfellow}, {Massana}, {Martinez-Delgado}, \&
  {Mu{\~n}oz}}]{choi_etal_2019}
{Choi} Y. {et~al.}, 2018, \apj, 869, 125

\bibitem[{{de Boer} {et~al}\mbox{.}(2018){de Boer}, {Belokurov}, {Koposov},
  {Ferrarese}, {Erkal}, {C{\^o}t{\'e}}, \& {Navarro}}]{deboer_gd1}
{de Boer} T.~J.~L., {Belokurov} V., {Koposov} S.~E., {Ferrarese} L., {Erkal}
  D., {C{\^o}t{\'e}} P., {Navarro} J.~F., 2018, \mnras, 477, 1893

\bibitem[{{Deason} {et~al}\mbox{.}(2017){Deason}, {Belokurov}, {Erkal},
  {Koposov}, \& {Mackey}}]{Deason2017}
{Deason} A.~J., {Belokurov} V., {Erkal} D., {Koposov} S.~E., {Mackey} D., 2017,
  \mnras, 467, 2636

\bibitem[{{Deason} {et~al}\mbox{.}(2012){Deason}, {Belokurov}, {Evans}, \&
  {An}}]{deason_etal_2012}
{Deason} A.~J., {Belokurov} V., {Evans} N.~W., {An} J., 2012, \mnras, 424, L44

\bibitem[{{Debattista} {et~al}\mbox{.}(2013){Debattista}, {Ro{\v s}kar},
  {Valluri}, {Quinn}, {Moore}, \& {Wadsley}}]{debattista_etal_2013}
{Debattista} V.~P., {Ro{\v s}kar} R., {Valluri} M., {Quinn} T., {Moore} B.,
  {Wadsley} J., 2013, \mnras, 434, 2971

\bibitem[{{Dey} {et~al}\mbox{.}(2018){Dey}, {Schlegel}, {Lang}, {Blum},
  {Burleigh}, {Fan}, {Findlay}, {Finkbeiner}, {Herrera}, {Juneau}, {Landriau},
  {Levi}, {McGreer}, {Meisner}, {Myers}, {Moustakas}, {Nugent}, {Patej},
  {Schlafly}, {Walker}, {Valdes}, {Weaver}, {Yeche}, {Zou}, {Zhou}, {Abareshi},
  {Abbott}, {Abolfathi}, {Aguilera}, {Allen}, {Alvarez}, {Annis}, {Aubert},
  {Bell}, {BenZvi}, {Bielby}, {Bolton}, {Briceno}, {Buckley- Geer}, {Butler},
  {Calamida}, {Carlberg}, {Carter}, {Casas}, {Castander}, {Choi}, {Comparat},
  {Cukanovaite}, {Delubac}, {DeVries}, {Dey}, {Dhungana}, {Dickinson}, {Ding},
  {Donaldson}, {Duan}, {Duckworth}, {Eftekharzadeh}, {Eisenstein}, {Etourneau},
  {Fagrelius}, {Farihi}, {Fitzpatrick}, {Font-Ribera}, {Fulmer}, {Gansicke},
  {Gaztanaga}, {George}, {Gerdes}, {Gontcho}, {Green}, {Guy}, {Harmer},
  {Hernandez}, {Honscheid}, {Lijuan}, {Huang}, {James}, {Jannuzi}, {Jiang},
  {Joyce}, {Karcher}, {Karkar}, {Kehoe}, {Kneib}, {Kueter-Young}, {Lan},
  {Lauer}, {Le Guillou}, {Le Van Suu}, {Lee}, {Lesser}, {Li}, {Mann},
  {Marshall}, {Mart{\'\i}nez-V{\'a}zquez}, {Martini}, {du Mas des Bourboux},
  {McManus}, {Menard}, {Metcalfe}, {Mu{\~n}oz-Guti{\'e}rrez}, {Najita},
  {Napier}, {Narayan}, {Newman}, {Nie}, {Nord}, {Norman}, {Olsen}, {Paat},
  {Palanque-Delabrouille}, {Peng}, {Poppett}, {Poremba}, {Prakash},
  {Rabinowitz}, {Raichoor}, {Rezaie}, {Robertson}, {Roe}, {Ross}, {Ross},
  {Rudnick}, {Safonova}, {Saha}, {Sanchez}, {Schweiker}, {Scott}, {Seo},
  {Shan}, {Silva}, {Soto}, {Sprayberry}, {Staten}, {Stillman}, {Stupak},
  {Summers}, {Sien Tie}, {Tirado}, {Vargas- Magana}, {Vivas}, {Wechsler},
  {Williams}, {Yang}, {Yang}, {Yapici}, {Zaritsky}, {Zenteno}, {Zhang},
  {Zhang}, {Zhou}, \& {Zhou}}]{DECALS2018}
{Dey} A. {et~al.}, 2018, arXiv e-prints, arXiv:1804.08657

\bibitem[{{Eadie} \& {Juri{\'c}}(2018)}]{eadie2018}
{Eadie} G., {Juri{\'c}} M., 2018, arXiv e-prints, arXiv:1810.10036

\bibitem[{{Erkal} \& {Belokurov}(2015)}]{Erkal2015}
{Erkal} D., {Belokurov} V., 2015, \mnras, 450, 1136

\bibitem[{{Erkal}, {Koposov} \& {Belokurov}(2017){Erkal}, {Koposov}, \&
  {Belokurov}}]{Erkal2017}
{Erkal} D., {Koposov} S.~E., {Belokurov} V., 2017, \mnras, 470, 60

\bibitem[{{Erkal} {et~al}\mbox{.}(2018){Erkal}, {Li}, {Koposov}, {Belokurov},
  {Balbinot}, {Bechtol}, {Buncher}, {Drlica-Wagner}, {Kuehn}, {Marshall},
  {Mart{\'{\i}}nez-V{\'a}zquez}, {Pace}, {Shipp}, {Simon}, {Stringer}, {Vivas},
  {Wechsler}, {Yanny}, {Abdalla}, {Allam}, {Annis}, {Avila}, {Bertin},
  {Brooks}, {Buckley-Geer}, {Burke}, {Carnero Rosell}, {Carrasco Kind},
  {Carretero}, {D'Andrea}, {da Costa}, {Davis}, {De Vicente}, {Doel}, {Eifler},
  {Evrard}, {Flaugher}, {Frieman}, {Garc{\'{\i}}a-Bellido}, {Gaztanaga},
  {Gerdes}, {Gruen}, {Gruendl}, {Gschwend}, {Gutierrez}, {Hartley},
  {Hollowood}, {Honscheid}, {James}, {Krause}, {Maia}, {March}, {Menanteau},
  {Miquel}, {Ogando}, {Plazas}, {Sanchez}, {Santiago}, {Scarpine}, {Schindler},
  {Sevilla-Noarbe}, {Smith}, {Smith}, {Soares-Santos}, {Sobreira}, {Suchyta},
  {Swanson}, {Tarle}, {Tucker}, \& {Walker}}]{tuciii_modelling}
{Erkal} D. {et~al.}, 2018, \mnras, 481, 3148

\bibitem[{{Erkal}, {Sanders} \& {Belokurov}(2016){Erkal}, {Sanders}, \&
  {Belokurov}}]{stray}
{Erkal} D., {Sanders} J.~L., {Belokurov} V., 2016, \mnras, 461, 1590

\bibitem[{{Errani}, {Pe{\~n}arrubia} \& {Walker}(2018){Errani},
  {Pe{\~n}arrubia}, \& {Walker}}]{errani_etal_2018}
{Errani} R., {Pe{\~n}arrubia} J., {Walker} M.~G., 2018, \mnras, 481, 5073

\bibitem[{{Evans}(1994)}]{evans1994}
{Evans} N.~W., 1994, \mnras, 267, 333

\bibitem[{{Fardal} {et~al}\mbox{.}(2018){Fardal}, {van der Marel}, {Sohn}, \&
  {del Pino Molina}}]{fardal2018}
{Fardal} M.~A., {van der Marel} R.~P., {Sohn} S.~T., {del Pino Molina} A.,
  2018, arXiv e-prints

\bibitem[{{Fardal} {et~al}\mbox{.}(2013){Fardal}, {Weinberg}, {Babul}, {Irwin},
  {Guhathakurta}, {Gilbert}, {Ferguson}, {Ibata}, {Lewis}, {Tanvir}, \&
  {Huxor}}]{Fardal2013}
{Fardal} M.~A. {et~al.}, 2013, \mnras, 434, 2779

\bibitem[{{Fellhauer} {et~al}\mbox{.}(2006){Fellhauer}, {Belokurov}, {Evans},
  {Wilkinson}, {Zucker}, {Gilmore}, {Irwin}, {Bramich}, {Vidrih}, {Wyse},
  {Beers}, \& {Brinkmann}}]{Fellhauer2006}
{Fellhauer} M. {et~al.}, 2006, \apj, 651, 167

\bibitem[{{Foreman-Mackey} {et~al}\mbox{.}(2013){Foreman-Mackey}, {Hogg},
  {Lang}, \& {Goodman}}]{emcee}
{Foreman-Mackey} D., {Hogg} D.~W., {Lang} D., {Goodman} J., 2013, \pasp, 125,
  306

\bibitem[{{Fritz} {et~al}\mbox{.}(2018){Fritz}, {Battaglia}, {Pawlowski},
  {Kallivayalil}, {van der Marel}, {Sohn}, {Brook}, \&
  {Besla}}]{fritz_pms_2018}
{Fritz} T.~K., {Battaglia} G., {Pawlowski} M.~S., {Kallivayalil} N., {van der
  Marel} R., {Sohn} S.~T., {Brook} C., {Besla} G., 2018, \aap, 619, A103

\bibitem[{{Fritz} \& {Kallivayalil}(2015)}]{pal5_pm}
{Fritz} T.~K., {Kallivayalil} N., 2015, \apj, 811, 123

\bibitem[{{Gaia Collaboration} {et~al}\mbox{.}(2018){Gaia Collaboration},
  {Brown}, {Vallenari}, {Prusti}, {de Bruijne}, {Babusiaux}, {Bailer-Jones},
  {Biermann}, {Evans}, {Eyer}, {Jansen}, {Jordi}, {Klioner}, {Lammers},
  {Lindegren}, {Luri}, {Mignard}, {Panem}, {Pourbaix}, {Randich}, {Sartoretti},
  {Siddiqui}, {Soubiran}, {van Leeuwen}, {Walton}, {Arenou}, {Bastian},
  {Cropper}, {Drimmel}, {Katz}, {Lattanzi}, {Bakker}, {Cacciari},
  {Casta{\~n}eda}, {Chaoul}, {Cheek}, {De Angeli}, {Fabricius}, {Guerra},
  {Holl}, {Masana}, {Messineo}, {Mowlavi}, {Nienartowicz}, {Panuzzo},
  {Portell}, {Riello}, {Seabroke}, {Tanga}, {Th{\'e}venin}, {Gracia-Abril},
  {Comoretto}, {Garcia-Reinaldos}, {Teyssier}, {Altmann}, {Andrae}, {Audard},
  {Bellas-Velidis}, {Benson}, {Berthier}, {Blomme}, {Burgess}, {Busso},
  {Carry}, {Cellino}, {Clementini}, {Clotet}, {Creevey}, {Davidson}, {De
  Ridder}, {Delchambre}, {Dell'Oro}, {Ducourant}, {Fern{\'a}ndez-
  Hern{\'a}ndez}, {Fouesneau}, {Fr{\'e}mat}, {Galluccio}, {Garc{\'\i}a-Torres},
  {Gonz{\'a}lez-N{\'u}{\~n}ez}, {Gonz{\'a}lez-Vidal}, {Gosset}, {Guy},
  {Halbwachs}, {Hambly}, {Harrison}, {Hern{\'a}ndez}, {Hestroffer}, {Hodgkin},
  {Hutton}, {Jasniewicz}, {Jean-Antoine-Piccolo}, {Jordan}, {Korn},
  {Krone-Martins}, {Lanzafame}, {Lebzelter}, {L{\"o}ffler}, {Manteiga},
  {Marrese}, {Mart{\'\i}n-Fleitas}, {Moitinho}, {Mora}, {Muinonen}, {Osinde},
  {Pancino}, {Pauwels}, {Petit}, {Recio-Blanco}, {Richards}, {Rimoldini},
  {Robin}, {Sarro}, {Siopis}, {Smith}, {Sozzetti}, {S{\"u}veges}, {Torra}, {van
  Reeven}, {Abbas}, {Abreu Aramburu}, {Accart}, {Aerts}, {Altavilla},
  {{\'A}lvarez}, {Alvarez}, {Alves}, {Anderson}, {Andrei}, {Anglada Varela},
  {Antiche}, {Antoja}, {Arcay}, {Astraatmadja}, {Bach}, {Baker},
  {Balaguer-N{\'u}{\~n}ez}, {Balm}, {Barache}, {Barata}, {Barbato}, {Barblan},
  {Barklem}, {Barrado}, {Barros}, {Barstow}, {Bartholom{\'e} Mu{\~n}oz},
  {Bassilana}, {Becciani}, {Bellazzini}, {Berihuete}, {Bertone}, {Bianchi},
  {Bienaym{\'e}}, {Blanco-Cuaresma}, {Boch}, {Boeche}, {Bombrun}, {Borrachero},
  {Bossini}, {Bouquillon}, {Bourda}, {Bragaglia}, {Bramante}, {Breddels},
  {Bressan}, {Brouillet}, {Br{\"u}semeister}, {Brugaletta}, {Bucciarelli},
  {Burlacu}, {Busonero}, {Butkevich}, {Buzzi}, {Caffau}, {Cancelliere},
  {Cannizzaro}, {Cantat-Gaudin}, {Carballo}, {Carlucci}, {Carrasco},
  {Casamiquela}, {Castellani}, {Castro-Ginard}, {Charlot}, {Chemin},
  {Chiavassa}, {Cocozza}, {Costigan}, {Cowell}, {Crifo}, {Crosta}, {Crowley},
  {Cuypers}, {Dafonte}, {Damerdji}, {Dapergolas}, {David}, {David}, {de
  Laverny}, {De Luise}, {De March}, {de Martino}, {de Souza}, {de Torres},
  {Debosscher}, {del Pozo}, {Delbo}, {Delgado}, {Delgado}, {Di Matteo},
  {Diakite}, {Diener}, {Distefano}, {Dolding}, {Drazinos}, {Dur{\'a}n},
  {Edvardsson}, {Enke}, {Eriksson}, {Esquej}, {Eynard Bontemps}, {Fabre},
  {Fabrizio}, {Faigler}, {Falc{\~a}o}, {Farr{\`a}s Casas}, {Federici},
  {Fedorets}, {Fernique}, {Figueras}, {Filippi}, {Findeisen}, {Fonti},
  {Fraile}, {Fraser}, {Fr{\'e}zouls}, {Gai}, {Galleti}, {Garabato},
  {Garc{\'\i}a-Sedano}, {Garofalo}, {Garralda}, {Gavel}, {Gavras}, {Gerssen},
  {Geyer}, {Giacobbe}, {Gilmore}, {Girona}, {Giuffrida}, {Glass}, {Gomes},
  {Granvik}, {Gueguen}, {Guerrier}, {Guiraud}, {Guti{\'e}rrez-S{\'a}nchez},
  {Haigron}, {Hatzidimitriou}, {Hauser}, {Haywood}, {Heiter}, {Helmi}, {Heu},
  {Hilger}, {Hobbs}, {Hofmann}, {Holland}, {Huckle}, {Hypki}, {Icardi},
  {Jan{\ss}en}, {Jevardat de Fombelle}, {Jonker}, {Juh{\'a}sz}, {Julbe},
  {Karampelas}, {Kewley}, {Klar}, {Kochoska}, {Kohley}, {Kolenberg},
  {Kontizas}, {Kontizas}, {Koposov}, {Kordopatis}, {Kostrzewa-Rutkowska},
  {Koubsky}, {Lambert}, {Lanza}, {Lasne}, {Lavigne}, {Le Fustec}, {Le
  Poncin-Lafitte}, {Lebreton}, {Leccia}, {Leclerc}, {Lecoeur-Taibi},
  {Lenhardt}, {Leroux}, {Liao}, {Licata}, {Lindstr{\o}m}, {Lister}, {Livanou},
  {Lobel}, {L{\'o}pez}, {Managau}, {Mann}, {Mantelet}, {Marchal}, {Marchant},
  {Marconi}, {Marinoni}, {Marschalk{\'o}}, {Marshall}, {Martino}, {Marton},
  {Mary}, {Massari}, {Matijevi{\v{c}}}, {Mazeh}, {McMillan}, {Messina},
  {Michalik}, {Millar}, {Molina}, {Molinaro}, {Moln{\'a}r}, {Montegriffo},
  {Mor}, {Morbidelli}, {Morel}, {Morris}, {Mulone}, {Muraveva}, {Musella},
  {Nelemans}, {Nicastro}, {Noval}, {O'Mullane}, {Ord{\'e}novic},
  {Ord{\'o}{\~n}ez-Blanco}, {Osborne}, {Pagani}, {Pagano}, {Pailler},
  {Palacin}, {Palaversa}, {Panahi}, {Pawlak}, {Piersimoni}, {Pineau}, {Plachy},
  {Plum}, {Poggio}, {Poujoulet}, {Pr{\v{s}}a}, {Pulone}, {Racero}, {Ragaini},
  {Rambaux}, {Ramos-Lerate}, {Regibo}, {Reyl{\'e}}, {Riclet}, {Ripepi}, {Riva},
  {Rivard}, {Rixon}, {Roegiers}, {Roelens}, {Romero-G{\'o}mez}, {Rowell},
  {Royer}, {Ruiz-Dern}, {Sadowski}, {Sagrist{\`a} Sell{\'e}s}, {Sahlmann},
  {Salgado}, {Salguero}, {Sanna}, {Santana- Ros}, {Sarasso}, {Savietto},
  {Schultheis}, {Sciacca}, {Segol}, {Segovia}, {S{\'e}gransan}, {Shih},
  {Siltala}, {Silva}, {Smart}, {Smith}, {Solano}, {Solitro}, {Sordo}, {Soria
  Nieto}, {Souchay}, {Spagna}, {Spoto}, {Stampa}, {Steele},
  {Steidelm{\"u}ller}, {Stephenson}, {Stoev}, {Suess}, {Surdej}, {Szabados},
  {Szegedi-Elek}, {Tapiador}, {Taris}, {Tauran}, {Taylor}, {Teixeira},
  {Terrett}, {Teyssandier}, {Thuillot}, {Titarenko}, {Torra Clotet}, {Turon},
  {Ulla}, {Utrilla}, {Uzzi}, {Vaillant}, {Valentini}, {Valette}, {van Elteren},
  {Van Hemelryck}, {van Leeuwen}, {Vaschetto}, {Vecchiato}, {Veljanoski},
  {Viala}, {Vicente}, {Vogt}, {von Essen}, {Voss}, {Votruba}, {Voutsinas},
  {Walmsley}, {Weiler}, {Wertz}, {Wevers}, {Wyrzykowski}, {Yoldas},
  {{\v{Z}}erjal}, {Ziaeepour}, {Zorec}, {Zschocke}, {Zucker}, {Zurbach}, \&
  {Zwitter}}]{Brown2018}
{Gaia Collaboration} {et~al.}, 2018, \aap, 616, A1

\bibitem[{{Garavito-Camargo} {et~al}\mbox{.}(2019){Garavito-Camargo}, {Besla},
  {Laporte}, {Johnston}, {G{\'o}mez}, \&
  {Watkins}}]{garavito_camargo_etal_2019}
{Garavito-Camargo} N., {Besla} G., {Laporte} C.~F.~P., {Johnston} K.~V.,
  {G{\'o}mez} F.~A., {Watkins} L.~L., 2019, arXiv e-prints

\bibitem[{{Gibbons}, {Belokurov} \& {Evans}(2014){Gibbons}, {Belokurov}, \&
  {Evans}}]{mLCS}
{Gibbons} S.~L.~J., {Belokurov} V., {Evans} N.~W., 2014, \mnras, 445, 3788

\bibitem[{{Gnedin} {et~al}\mbox{.}(2010){Gnedin}, {Brown}, {Geller}, \&
  {Kenyon}}]{gnedin_etal_2010}
{Gnedin} O.~Y., {Brown} W.~R., {Geller} M.~J., {Kenyon} S.~J., 2010, \apjl,
  720, L108

\bibitem[{{Gnedin} {et~al}\mbox{.}(2005){Gnedin}, {Gould},
  {Miralda-Escud{\'e}}, \& {Zentner}}]{gnedin_hvs}
{Gnedin} O.~Y., {Gould} A., {Miralda-Escud{\'e}} J., {Zentner} A.~R., 2005,
  \apj, 634, 344

\bibitem[{{G{\'o}mez} {et~al}\mbox{.}(2015){G{\'o}mez}, {Besla}, {Carpintero},
  {Villalobos}, {O'Shea}, \& {Bell}}]{gomez_etal_2015}
{G{\'o}mez} F.~A., {Besla} G., {Carpintero} D.~D., {Villalobos} {\'A}.,
  {O'Shea} B.~W., {Bell} E.~F., 2015, \apj, 802, 128

\bibitem[{{G{\'o}mez} {et~al}\mbox{.}(2016){G{\'o}mez}, {White}, {Marinacci},
  {Slater}, {Grand}, {Springel}, \& {Pakmor}}]{gomez_etal_2016}
{G{\'o}mez} F.~A., {White} S.~D.~M., {Marinacci} F., {Slater} C.~T., {Grand}
  R.~J.~J., {Springel} V., {Pakmor} R., 2016, \mnras, 456, 2779

\bibitem[{{Gravity Collaboration} {et~al}\mbox{.}(2018){Gravity Collaboration},
  {Abuter}, {Amorim}, {Anugu}, {Baub{\"o}ck}, {Benisty}, {Berger}, {Blind},
  {Bonnet}, {Brandner}, {Buron}, {Collin}, {Chapron}, {Cl{\'e}net}, {Coud{\'e}
  Du Foresto}, {de Zeeuw}, {Deen}, {Delplancke-Str{\"o}bele}, {Dembet},
  {Dexter}, {Duvert}, {Eckart}, {Eisenhauer}, {Finger}, {F{\"o}rster
  Schreiber}, {F{\'e}dou}, {Garcia}, {Garcia Lopez}, {Gao}, {Gendron},
  {Genzel}, {Gillessen}, {Gordo}, {Habibi}, {Haubois}, {Haug}, {Hau{\ss}mann},
  {Henning}, {Hippler}, {Horrobin}, {Hubert}, {Hubin}, {Jimenez Rosales},
  {Jochum}, {Jocou}, {Kaufer}, {Kellner}, {Kendrew}, {Kervella}, {Kok},
  {Kulas}, {Lacour}, {Lapeyr{\`e}re}, {Lazareff}, {Le Bouquin}, {L{\'e}na},
  {Lippa}, {Lenzen}, {M{\'e}rand}, {M{\"u}ler}, {Neumann}, {Ott}, {Palanca},
  {Paumard}, {Pasquini}, {Perraut}, {Perrin}, {Pfuhl}, {Plewa}, {Rabien},
  {Ram{\'{\i}}rez}, {Ramos}, {Rau}, {Rodr{\'{\i}}guez-Coira}, {Rohloff},
  {Rousset}, {Sanchez-Bermudez}, {Scheithauer}, {Sch{\"o}ller}, {Schuler},
  {Spyromilio}, {Straub}, {Straubmeier}, {Sturm}, {Tacconi}, {Tristram},
  {Vincent}, {von Fellenberg}, {Wank}, {Waisberg}, {Widmann}, {Wieprecht},
  {Wiest}, {Wiezorrek}, {Woillez}, {Yazici}, {Ziegler}, \& {Zins}}]{s2_GR_R0}
{Gravity Collaboration} {et~al.}, 2018, \aap, 615, L15

\bibitem[{{Grillmair}(2006)}]{orphan_disc_g}
{Grillmair} C.~J., 2006, \apjl, 645, L37

\bibitem[{{Grillmair} \& {Dionatos}(2006)}]{gd1_disc}
{Grillmair} C.~J., {Dionatos} O., 2006, \apjl, 643, L17

\bibitem[{{Hayes}, {Law} \& {Majewski}(2018){Hayes}, {Law}, \&
  {Majewski}}]{hayes_sgr_lsr}
{Hayes} C.~R., {Law} D.~R., {Majewski} S.~R., 2018, \apjl, 867, L20

\bibitem[{{Helmi}(2004)}]{helmi04}
{Helmi} A., 2004, \apjl, 610, L97

\bibitem[{{Hendel} {et~al}\mbox{.}(2018){Hendel}, {Scowcroft}, {Johnston},
  {Fardal}, {van der Marel}, {Sohn}, {Price-Whelan}, {Beaton}, {Besla}, {Bono},
  {Cioni}, {Clementini}, {Cohen}, {Fabrizio}, {Freedman}, {Garofalo},
  {Grillmair}, {Kallivayalil}, {Kollmeier}, {Law}, {Madore}, {Majewski},
  {Marengo}, {Monson}, {Neeley}, {Nidever}, {Pietrzy{\'n}ski}, {Seibert},
  {Sesar}, {Smith}, {Soszy{\'n}ski}, \& {Udalski}}]{hendel_etal_2018}
{Hendel} D. {et~al.}, 2018, \mnras, 479, 570

\bibitem[{{Hernquist}(1990)}]{hernquist_profile}
{Hernquist} L., 1990, \apj, 356, 359

\bibitem[{{Hunter}(2007)}]{matplotlib}
{Hunter} J.~D., 2007, Computing in Science Engineering, 9, 90

\bibitem[{{Ibata} {et~al}\mbox{.}(2004){Ibata}, {Chapman}, {Ferguson}, {Irwin},
  {Lewis}, \& {McConnachie}}]{Ibata2004}
{Ibata} R., {Chapman} S., {Ferguson} A.~M.~N., {Irwin} M., {Lewis} G.,
  {McConnachie} A., 2004, \mnras, 351, 117

\bibitem[{{Ibata} {et~al}\mbox{.}(2001){Ibata}, {Lewis}, {Irwin}, {Totten}, \&
  {Quinn}}]{Ibata2001}
{Ibata} R., {Lewis} G.~F., {Irwin} M., {Totten} E., {Quinn} T., 2001, \apj,
  551, 294

\bibitem[{{Ibata} {et~al}\mbox{.}(2002){Ibata}, {Lewis}, {Irwin}, \&
  {Quinn}}]{Ibata2002}
{Ibata} R.~A., {Lewis} G.~F., {Irwin} M.~J., {Quinn} T., 2002, \mnras, 332, 915

\bibitem[{{Jethwa}, {Erkal} \& {Belokurov}(2016){Jethwa}, {Erkal}, \&
  {Belokurov}}]{jethwa_lmc_sats}
{Jethwa} P., {Erkal} D., {Belokurov} V., 2016, \mnras, 461, 2212

\bibitem[{{Johnston}, {Law} \& {Majewski}(2005){Johnston}, {Law}, \&
  {Majewski}}]{Johnston2005}
{Johnston} K.~V., {Law} D.~R., {Majewski} S.~R., 2005, \apj, 619, 800

\bibitem[{{Johnston}, {Spergel} \& {Haydn}(2002){Johnston}, {Spergel}, \&
  {Haydn}}]{Johnston2002}
{Johnston} K.~V., {Spergel} D.~N., {Haydn} C., 2002, \apj, 570, 656

\bibitem[{{Johnston} {et~al}\mbox{.}(1999){Johnston}, {Zhao}, {Spergel}, \&
  {Hernquist}}]{johnston99}
{Johnston} K.~V., {Zhao} H., {Spergel} D.~N., {Hernquist} L., 1999, \apjl, 512,
  L109

\bibitem[{Jones {et~al}\mbox{.}(2001--)Jones, Oliphant, Peterson,
  {et~al.}}]{scipy}
Jones E., Oliphant T., Peterson P., {et~al.}, 2001--, {SciPy}: Open source
  scientific tools for {Python}.
  \href{http://www.scipy.org/}{http://www.scipy.org/}

\bibitem[{{Kallivayalil} {et~al}\mbox{.}(2018){Kallivayalil}, {Sales},
  {Zivick}, {Fritz}, {Del Pino}, {Sohn}, {Besla}, {van der Marel}, {Navarro},
  \& {Sacchi}}]{kallivayalil_lmc_sats}
{Kallivayalil} N. {et~al.}, 2018, \apj, 867, 19

\bibitem[{{Kallivayalil} {et~al}\mbox{.}(2013){Kallivayalil}, {van der Marel},
  {Besla}, {Anderson}, \& {Alcock}}]{lmc_pms}
{Kallivayalil} N., {van der Marel} R.~P., {Besla} G., {Anderson} J., {Alcock}
  C., 2013, \apj, 764, 161

\bibitem[{{Kleyna} {et~al}\mbox{.}(2005){Kleyna}, {Wilkinson}, {Evans}, \&
  {Gilmore}}]{kleyna_etal_2005}
{Kleyna} J.~T., {Wilkinson} M.~I., {Evans} N.~W., {Gilmore} G., 2005, \apjl,
  630, L141

\bibitem[{{Kochanek}, {Keeton} \& {McLeod}(2001){Kochanek}, {Keeton}, \&
  {McLeod}}]{Kochanek2001}
{Kochanek} C.~S., {Keeton} C.~R., {McLeod} B.~A., 2001, \apj, 547, 50

\bibitem[{{Koposov} {et~al}\mbox{.}(2019){Koposov}, {Belokurov}, {Li}, {Mateu},
  {Erkal}, {Grillmair}, {Hendel}, {Price-Whelan}, {Laporte}, {Hawkins}, {Sohn},
  {del Pino}, {Evans}, {Slater}, {Kallivayalil}, \&
  {Navarro}}]{koposov_etal_orphan}
{Koposov} S.~E. {et~al.}, 2019, \mnras, 485, 4726

\bibitem[{{Koposov}, {Rix} \& {Hogg}(2010){Koposov}, {Rix}, \&
  {Hogg}}]{koposov_gd1}
{Koposov} S.~E., {Rix} H.-W., {Hogg} D.~W., 2010, \apj, 712, 260

\bibitem[{{K{\"u}pper} {et~al}\mbox{.}(2015){K{\"u}pper}, {Balbinot}, {Bonaca},
  {Johnston}, {Hogg}, {Kroupa}, \& {Santiago}}]{Kupper2015}
{K{\"u}pper} A.~H.~W., {Balbinot} E., {Bonaca} A., {Johnston} K.~V., {Hogg}
  D.~W., {Kroupa} P., {Santiago} B.~X., 2015, \apj, 803, 80

\bibitem[{{Laporte} {et~al}\mbox{.}(2018){Laporte}, {G{\'o}mez}, {Besla},
  {Johnston}, \& {Garavito-Camargo}}]{laporte2018}
{Laporte} C.~F.~P., {G{\'o}mez} F.~A., {Besla} G., {Johnston} K.~V.,
  {Garavito-Camargo} N., 2018, \mnras, 473, 1218

\bibitem[{{Law} \& {Majewski}(2010)}]{law_majewski_2010}
{Law} D.~R., {Majewski} S.~R., 2010, \apj, 714, 229

\bibitem[{{Levine}, {Blitz} \& {Heiles}(2006){Levine}, {Blitz}, \&
  {Heiles}}]{levine06}
{Levine} E.~S., {Blitz} L., {Heiles} C., 2006, \apj, 643, 881

\bibitem[{{Mackey} {et~al}\mbox{.}(2016){Mackey}, {Koposov}, {Erkal},
  {Belokurov}, {Da Costa}, \& {G{\'o}mez}}]{mackey_lmc_disk}
{Mackey} A.~D., {Koposov} S.~E., {Erkal} D., {Belokurov} V., {Da Costa} G.~S.,
  {G{\'o}mez} F.~A., 2016, \mnras, 459, 239

\bibitem[{{Mackey} {et~al}\mbox{.}(2018){Mackey}, {Koposov}, {Da Costa},
  {Belokurov}, {Erkal}, \& {Kuzma}}]{mackey_2018}
{Mackey} D., {Koposov} S., {Da Costa} G., {Belokurov} V., {Erkal} D., {Kuzma}
  P., 2018, \apjl, 858, L21

\bibitem[{{Majewski} {et~al}\mbox{.}(2006){Majewski}, {Law}, {Polak}, \&
  {Patterson}}]{majewski_r0lsr_2006}
{Majewski} S.~R., {Law} D.~R., {Polak} A.~A., {Patterson} R.~J., 2006, \apjl,
  637, L25

\bibitem[{{Malhan} \& {Ibata}(2017)}]{malhan_ibata_reflex}
{Malhan} K., {Ibata} R.~A., 2017, \mnras, 471, 1005

\bibitem[{{Malhan} \& {Ibata}(2018)}]{malhan_stream_finder}
{Malhan} K., {Ibata} R.~A., 2018, \mnras, 477, 4063

\bibitem[{{Mandelbaum} {et~al}\mbox{.}(2006){Mandelbaum}, {Seljak},
  {Kauffmann}, {Hirata}, \& {Brinkmann}}]{Mandelbaum2006}
{Mandelbaum} R., {Seljak} U., {Kauffmann} G., {Hirata} C.~M., {Brinkmann} J.,
  2006, \mnras, 368, 715

\bibitem[{{Miyamoto} \& {Nagai}(1975)}]{mn_disk}
{Miyamoto} M., {Nagai} R., 1975, \pasj, 27, 533

\bibitem[{{Moster}, {Naab} \& {White}(2013){Moster}, {Naab}, \&
  {White}}]{Moster2013}
{Moster} B.~P., {Naab} T., {White} S.~D.~M., 2013, \mnras, 428, 3121

\bibitem[{{Navarrete} {et~al}\mbox{.}(2019){Navarrete}, {Belokurov}, {Catelan},
  {Jethwa}, {Koposov}, {Carballo-Bello}, {Jofr{\'e}}, {Erkal}, {Duffau}, \&
  {Corral-Santana}}]{Navarrete2018}
{Navarrete} C. {et~al.}, 2019, \mnras, 483, 4160

\bibitem[{{Navarro}, {Frenk} \& {White}(1997){Navarro}, {Frenk}, \&
  {White}}]{nfw_1997}
{Navarro} J.~F., {Frenk} C.~S., {White} S.~D.~M., 1997, \apj, 490, 493

\bibitem[{{Newberg} {et~al}\mbox{.}(2010){Newberg}, {Willett}, {Yanny}, \&
  {Xu}}]{newberg_etal_2010}
{Newberg} H.~J., {Willett} B.~A., {Yanny} B., {Xu} Y., 2010, \apj, 711, 32

\bibitem[{{Ngan} \& {Carlberg}(2014)}]{ngan_carlberg}
{Ngan} W.~H.~W., {Carlberg} R.~G., 2014, \apj, 788, 181

\bibitem[{{Nidever} {et~al}\mbox{.}(2019){Nidever}, {Olsen}, {Choi}, {de Boer},
  {Blum}, {Bell}, {Zaritsky}, {Martin}, {Saha}, {Conn}, {Besla}, {van der
  Marel}, {No{\"e}l}, {Monachesi}, {Stringfellow}, {Massana}, {Cioni},
  {Gallart}, {Monelli}, {Martinez-Delgado}, {Mu{\~n}oz}, {Majewski}, {Vivas},
  {Walker}, {Kaleida}, \& {Chu}}]{nidever_etal_2019}
{Nidever} D.~L. {et~al.}, 2019, \apj, 874, 118

\bibitem[{{Odenkirchen} {et~al}\mbox{.}(2001){Odenkirchen}, {Grebel},
  {Rockosi}, {Dehnen}, {Ibata}, {Rix}, {Stolte}, {Wolf}, {Anderson}, {Bahcall},
  {Brinkmann}, {Csabai}, {Hennessy}, {Hindsley}, {Ivezi{\'c}}, {Lupton},
  {Munn}, {Pier}, {Stoughton}, \& {York}}]{pal5_disc}
{Odenkirchen} M. {et~al.}, 2001, \apjl, 548, L165

\bibitem[{{Pe{\~n}arrubia} {et~al}\mbox{.}(2016){Pe{\~n}arrubia}, {G{\'o}mez},
  {Besla}, {Erkal}, \& {Ma}}]{penarrubia_lmc_mass}
{Pe{\~n}arrubia} J., {G{\'o}mez} F.~A., {Besla} G., {Erkal} D., {Ma} Y.-Z.,
  2016, \mnras, 456, L54

\bibitem[{{Pearson} {et~al}\mbox{.}(2015){Pearson}, {K{\"u}pper}, {Johnston},
  \& {Price-Whelan}}]{Pearson2015}
{Pearson} S., {K{\"u}pper} A.~H.~W., {Johnston} K.~V., {Price-Whelan} A.~M.,
  2015, \apj, 799, 28

\bibitem[{{Perez} \& {Granger}(2007)}]{IPython}
{Perez} F., {Granger} B.~E., 2007, Computing in Science Engineering, 9, 21

\bibitem[{{Pietrzy{\'n}ski} {et~al}\mbox{.}(2013){Pietrzy{\'n}ski}, {Graczyk},
  {Gieren}, {Thompson}, {Pilecki}, {Udalski}, {Soszy{\'n}ski}, {Koz{\l}owski},
  {Konorski}, {Suchomska}, {Bono}, {Moroni}, {Villanova}, {Nardetto},
  {Bresolin}, {Kudritzki}, {Storm}, {Gallenne}, {Smolec}, {Minniti}, {Kubiak},
  {Szyma{\'n}ski}, {Poleski}, {Wyrzykowski}, {Ulaczyk}, {Pietrukowicz},
  {G{\'o}rski}, \& {Karczmarek}}]{lmc_dist}
{Pietrzy{\'n}ski} G. {et~al.}, 2013, \nat, 495, 76

\bibitem[{{Price-Whelan} \& {Bonaca}(2018)}]{PW2018}
{Price-Whelan} A.~M., {Bonaca} A., 2018, \apj, 863, L20

\bibitem[{{Price-Whelan} {et~al}\mbox{.}(2018){Price-Whelan}, {Sip{\H{o}}cz},
  {G{\"u}nther}, {Lim}, {Crawford}, {Conseil}, {Shupe}, {Craig}, {Dencheva},
  {Ginsburg}, {VanderPlas}, {Bradley}, {P{\'e}rez-Su{\'a}rez}, {de Val-Borro},
  {Paper Contributors}, {Aldcroft}, {Cruz}, {Robitaille}, {Tollerud},
  {Coordination Committee}, {Ardelean}, {Babej}, {Bach}, {Bachetti}, {Bakanov},
  {Bamford}, {Barentsen}, {Barmby}, {Baumbach}, {Berry}, {Biscani}, {Boquien},
  {Bostroem}, {Bouma}, {Brammer}, {Bray}, {Breytenbach}, {Buddelmeijer},
  {Burke}, {Calderone}, {Cano Rodr{\'\i}guez}, {Cara}, {Cardoso}, {Cheedella},
  {Copin}, {Corrales}, {Crichton}, {D{\textquoteright}Avella}, {Deil},
  {Depagne}, {Dietrich}, {Donath}, {Droettboom}, {Earl}, {Erben}, {Fabbro},
  {Ferreira}, {Finethy}, {Fox}, {Garrison}, {Gibbons}, {Goldstein}, {Gommers},
  {Greco}, {Greenfield}, {Groener}, {Grollier}, {Hagen}, {Hirst}, {Homeier},
  {Horton}, {Hosseinzadeh}, {Hu}, {Hunkeler}, {Ivezi{\'c}}, {Jain}, {Jenness},
  {Kanarek}, {Kendrew}, {Kern}, {Kerzendorf}, {Khvalko}, {King}, {Kirkby},
  {Kulkarni}, {Kumar}, {Lee}, {Lenz}, {Littlefair}, {Ma}, {Macleod},
  {Mastropietro}, {McCully}, {Montagnac}, {Morris}, {Mueller}, {Mumford},
  {Muna}, {Murphy}, {Nelson}, {Nguyen}, {Ninan}, {N{\"o}the}, {Ogaz}, {Oh},
  {Parejko}, {Parley}, {Pascual}, {Patil}, {Patil}, {Plunkett}, {Prochaska},
  {Rastogi}, {Reddy Janga}, {Sabater}, {Sakurikar}, {Seifert}, {Sherbert},
  {Sherwood-Taylor}, {Shih}, {Sick}, {Silbiger}, {Singanamalla}, {Singer},
  {Sladen}, {Sooley}, {Sornarajah}, {Streicher}, {Teuben}, {Thomas},
  {Tremblay}, {Turner}, {Terr{\'o}n}, {van Kerkwijk}, {de la Vega}, {Watkins},
  {Weaver}, {Whitmore}, {Woillez}, {Zabalza}, \& {Contributors}}]{astropy:2018}
{Price-Whelan} A.~M. {et~al.}, 2018, \aj, 156, 123

\bibitem[{{Putman} {et~al}\mbox{.}(2003){Putman}, {Staveley-Smith}, {Freeman},
  {Gibson}, \& {Barnes}}]{putman_magellanic_stream}
{Putman} M.~E., {Staveley-Smith} L., {Freeman} K.~C., {Gibson} B.~K., {Barnes}
  D.~G., 2003, \apj, 586, 170

\bibitem[{{Reid} \& {Brunthaler}(2004)}]{pm_sgr_a}
{Reid} M.~J., {Brunthaler} A., 2004, \apj, 616, 872

\bibitem[{{Richstone}(1980)}]{richstone1980}
{Richstone} D.~O., 1980, \apj, 238, 103

\bibitem[{{Sales} {et~al}\mbox{.}(2011){Sales}, {Navarro}, {Cooper}, {White},
  {Frenk}, \& {Helmi}}]{sales2011}
{Sales} L.~V., {Navarro} J.~F., {Cooper} A.~P., {White} S.~D.~M., {Frenk}
  C.~S., {Helmi} A., 2011, \mnras, 418, 648

\bibitem[{{Sand}, {Treu} \& {Ellis}(2002){Sand}, {Treu}, \& {Ellis}}]{sand02}
{Sand} D.~J., {Treu} T., {Ellis} R.~S., 2002, \apjl, 574, L129

\bibitem[{{Sanders} \& {Binney}(2013)}]{sanders_binney_2013}
{Sanders} J.~L., {Binney} J., 2013, \mnras, 433, 1813

\bibitem[{{Schommer} {et~al}\mbox{.}(1992){Schommer}, {Suntzeff}, {Olszewski},
  \& {Harris}}]{schommer_etal_1992}
{Schommer} R.~A., {Suntzeff} N.~B., {Olszewski} E.~W., {Harris} H.~C., 1992,
  \aj, 103, 447

\bibitem[{{Sch{\"o}nrich}, {Binney} \& {Dehnen}(2010){Sch{\"o}nrich}, {Binney},
  \& {Dehnen}}]{schonrich_vlsr}
{Sch{\"o}nrich} R., {Binney} J., {Dehnen} W., 2010, \mnras, 403, 1829

\bibitem[{{Shipp} {et~al}\mbox{.}(2018){Shipp}, {Drlica-Wagner}, {Balbinot},
  {Ferguson}, {Erkal}, {Li}, {Bechtol}, {Belokurov}, {Buncher}, {Carollo},
  {Carrasco Kind}, {Kuehn}, {Marshall}, {Pace}, {Rykoff}, {Sevilla-Noarbe},
  {Sheldon}, {Strigari}, {Vivas}, {Yanny}, {Zenteno}, {Abbott}, {Abdalla},
  {Allam}, {Avila}, {Bertin}, {Brooks}, {Burke}, {Carretero}, {Castander},
  {Cawthon}, {Crocce}, {Cunha}, {D'Andrea}, {da Costa}, {Davis}, {De Vicente},
  {Desai}, {Diehl}, {Doel}, {Evrard}, {Flaugher}, {Fosalba}, {Frieman},
  {Garc{\'{\i}}a-Bellido}, {Gaztanaga}, {Gerdes}, {Gruen}, {Gruendl},
  {Gschwend}, {Gutierrez}, {Hartley}, {Honscheid}, {Hoyle}, {James}, {Johnson},
  {Krause}, {Kuropatkin}, {Lahav}, {Lin}, {Maia}, {March}, {Martini},
  {Menanteau}, {Miller}, {Miquel}, {Nichol}, {Plazas}, {Romer}, {Sako},
  {Sanchez}, {Santiago}, {Scarpine}, {Schindler}, {Schubnell}, {Smith},
  {Smith}, {Sobreira}, {Suchyta}, {Swanson}, {Tarle}, {Thomas}, {Tucker},
  {Walker}, {Wechsler}, \& {DES Collaboration}}]{des_streams}
{Shipp} N. {et~al.}, 2018, \apj, 862, 114

\bibitem[{{Siegal-Gaskins} \& {Valluri}(2008)}]{Siegal2008}
{Siegal-Gaskins} J.~M., {Valluri} M., 2008, \apj, 681, 40

\bibitem[{{Simon}(2018)}]{simon_2018}
{Simon} J.~D., 2018, \apj, 863, 89

\bibitem[{{Stanimirovi{\'c}}, {Staveley-Smith} \&
  {Jones}(2004){Stanimirovi{\'c}}, {Staveley-Smith}, \&
  {Jones}}]{stanimirovic_etal_2004}
{Stanimirovi{\'c}} S., {Staveley-Smith} L., {Jones} P.~A., 2004, \apj, 604, 176

\bibitem[{{Treu} \& {Koopmans}(2004)}]{Treu2004}
{Treu} T., {Koopmans} L.~V.~E., 2004, \apj, 611, 739

\bibitem[{{van der Marel}(2006)}]{vandermarel_2006}
{van der Marel} R.~P., 2006, in The Local Group as an Astrophysical Laboratory,
  {Livio} M., {Brown} T.~M., eds., Vol.~17, pp. 47--71

\bibitem[{{van der Marel} {et~al}\mbox{.}(2002){van der Marel}, {Alves},
  {Hardy}, \& {Suntzeff}}]{lmc_vr}
{van der Marel} R.~P., {Alves} D.~R., {Hardy} E., {Suntzeff} N.~B., 2002, \aj,
  124, 2639

\bibitem[{{van der Marel} \&
  {Kallivayalil}(2014)}]{vandermarel_kallivayalil_2014}
{van der Marel} R.~P., {Kallivayalil} N., 2014, \apj, 781, 121

\bibitem[{{van der Walt}, {Colbert} \& {Varoquaux}(2011){van der Walt},
  {Colbert}, \& {Varoquaux}}]{numpy}
{van der Walt} S., {Colbert} S.~C., {Varoquaux} G., 2011, Computing in Science
  Engineering, 13, 22

\bibitem[{{Vasiliev}(2019)}]{vasiliev_2018}
{Vasiliev} E., 2019, \mnras, 484, 2832

\bibitem[{{Vera-Ciro} \& {Helmi}(2013)}]{vera-ciro_helmi_2013}
{Vera-Ciro} C., {Helmi} A., 2013, \apjl, 773, L4

\bibitem[{{Vesperini} \& {Weinberg}(2000)}]{vesperini_weinberg_2000}
{Vesperini} E., {Weinberg} M.~D., 2000, \apj, 534, 598

\bibitem[{{Walker} {et~al}\mbox{.}(2009){Walker}, {Mateo}, {Olszewski},
  {Pe{\~n}arrubia}, {Evans}, \& {Gilmore}}]{walker_etal_2009}
{Walker} M.~G., {Mateo} M., {Olszewski} E.~W., {Pe{\~n}arrubia} J., {Evans}
  N.~W., {Gilmore} G., 2009, \apj, 704, 1274

\bibitem[{{Watkins}, {Evans} \& {An}(2010){Watkins}, {Evans}, \&
  {An}}]{watkins_etal_2010}
{Watkins} L.~L., {Evans} N.~W., {An} J.~H., 2010, \mnras, 406, 264

\bibitem[{{Watkins} {et~al}\mbox{.}(2018){Watkins}, {van der Marel}, {Sohn}, \&
  {Evans}}]{watkins_etal_2018}
{Watkins} L.~L., {van der Marel} R.~P., {Sohn} S.~T., {Evans} N.~W., 2018,
  arXiv e-prints, arXiv:1804.11348

\bibitem[{{Weinberg}(1989)}]{weinberg89}
{Weinberg} M.~D., 1989, \mnras, 239, 549

\bibitem[{{Weinberg}(1998)}]{weinberg98}
{Weinberg} M.~D., 1998, \mnras, 299, 499

\bibitem[{{Weinberg} \& {Blitz}(2006)}]{weinberg06}
{Weinberg} M.~D., {Blitz} L., 2006, \apjl, 641, L33

\bibitem[{{White}(2001)}]{White2001}
{White} M., 2001, \aap, 367, 27

\bibitem[{{Wilkinson} \& {Evans}(1999)}]{Wilkinson1999}
{Wilkinson} M.~I., {Evans} N.~W., 1999, \mnras, 310, 645

\bibitem[{{Xue} {et~al}\mbox{.}(2008){Xue}, {Rix}, {Zhao}, {Re Fiorentin},
  {Naab}, {Steinmetz}, {van den Bosch}, {Beers}, {Lee}, {Bell}, {Rockosi},
  {Yanny}, {Newberg}, {Wilhelm}, {Kang}, {Smith}, \& {Schneider}}]{Xue2008}
{Xue} X.~X. {et~al.}, 2008, \apj, 684, 1143

\bibitem[{{Yoon}, {Johnston} \& {Hogg}(2011){Yoon}, {Johnston}, \&
  {Hogg}}]{Yoon2011}
{Yoon} J.~H., {Johnston} K.~V., {Hogg} D.~W., 2011, \apj, 731, 58

\bibitem[{{Zhu} {et~al}\mbox{.}(2016){Zhu}, {Marinacci}, {Maji}, {Li},
  {Springel}, \& {Hernquist}}]{zhu_etal_2016}
{Zhu} Q., {Marinacci} F., {Maji} M., {Li} Y., {Springel} V., {Hernquist} L.,
  2016, \mnras, 458, 1559

\bibitem[{{Zivick} {et~al}\mbox{.}(2018){Zivick}, {Kallivayalil}, {van der
  Marel}, {Besla}, {Linden}, {Koz{\l}owski}, {Fritz}, {Kochanek}, {Anderson},
  {Sohn}, {Geha}, \& {Alcock}}]{zivick_etal_2018}
{Zivick} P. {et~al.}, 2018, \apj, 864, 55

\end{thebibliography}
